\documentclass[aps,pra,twocolumn,showpacs,superscriptaddress]{revtex4-1}

\usepackage{graphicx}
\usepackage{dcolumn}
\usepackage{color}
\usepackage{array,multirow,makecell}
\usepackage{bm}
\usepackage{nicefrac}

\begin{document}
\title{Three-dimensional critical phase diagram of the Ising antiferromagnet CeRh$_2$Si$_2$\\ under intense magnetic field and pressure}

\author{W. Knafo}
\affiliation{Laboratoire National des Champs Magn\'etiques Intenses, UPR 3228, CNRS-UPS-INSA-UGA, 143 Avenue de Rangueil, 31400 Toulouse, France}

\author{R. Settai}
\affiliation{Department of Physics, Faculty of Science, Niigata University,  8050 Ninocho Ikarashi, Nishi-ku, Niigata 950-2181, Japan}

\author{D. Braithwaite}
\affiliation{Universit\'{e} Grenoble Alpes and CEA, INAC-PHELIQS, F-38000 Grenoble, France}

\author{S. Kurahashi}
\affiliation{Graduate School of Science and Technology, Niigata University, 8050 Ninocho Ikarashi, Nishi-ku, Niigata 950-2181, Japan}

\author{D. Aoki}
\affiliation{Universit\'{e} Grenoble Alpes and CEA, INAC-PHELIQS, F-38000 Grenoble, France}
\affiliation{Institute for Materials Research, Tohoku University, Ibaraki 311-1313, Japan}

\author{J. Flouquet}
\affiliation{Universit\'{e} Grenoble Alpes and CEA, INAC-PHELIQS, F-38000 Grenoble, France}

\pacs{71.27.+a,74.70.Tx,75.30.Kz,75.30.Mb}

\date{\today}

\begin{abstract}
Using novel instrumentation to combine extreme conditions of intense pulsed magnetic field up to 60~T and high pressure up to 4~GPa, we have established the three-dimensional (3D) magnetic field - pressure - temperature phase diagram of a pure stoichiometric heavy-fermion antiferromagnet (CeRh$_2$Si$_2$). We find a temperature- and pressure-dependent decoupling of the critical and pseudo-metamagnetic fields, at the borderlines of antiferromagnetism and strongly-correlated paramagnetism. This 3D phase diagram is representative of a class of heavy-fermion Ising antiferromagnets, where long-range magnetic ordering is decoupled from a maximum in the magnetic susceptibility. The combination of extreme conditions enabled us to characterize different quantum phase transitions, where peculiar quantum critical properties are revealed. The interest to couple the effects of magnetic field and pressure on quantum-critical correlated-electron systems is stressed.\end{abstract}

\maketitle

\section{Introduction}

\begin{figure}
\includegraphics[width=1.02\columnwidth]{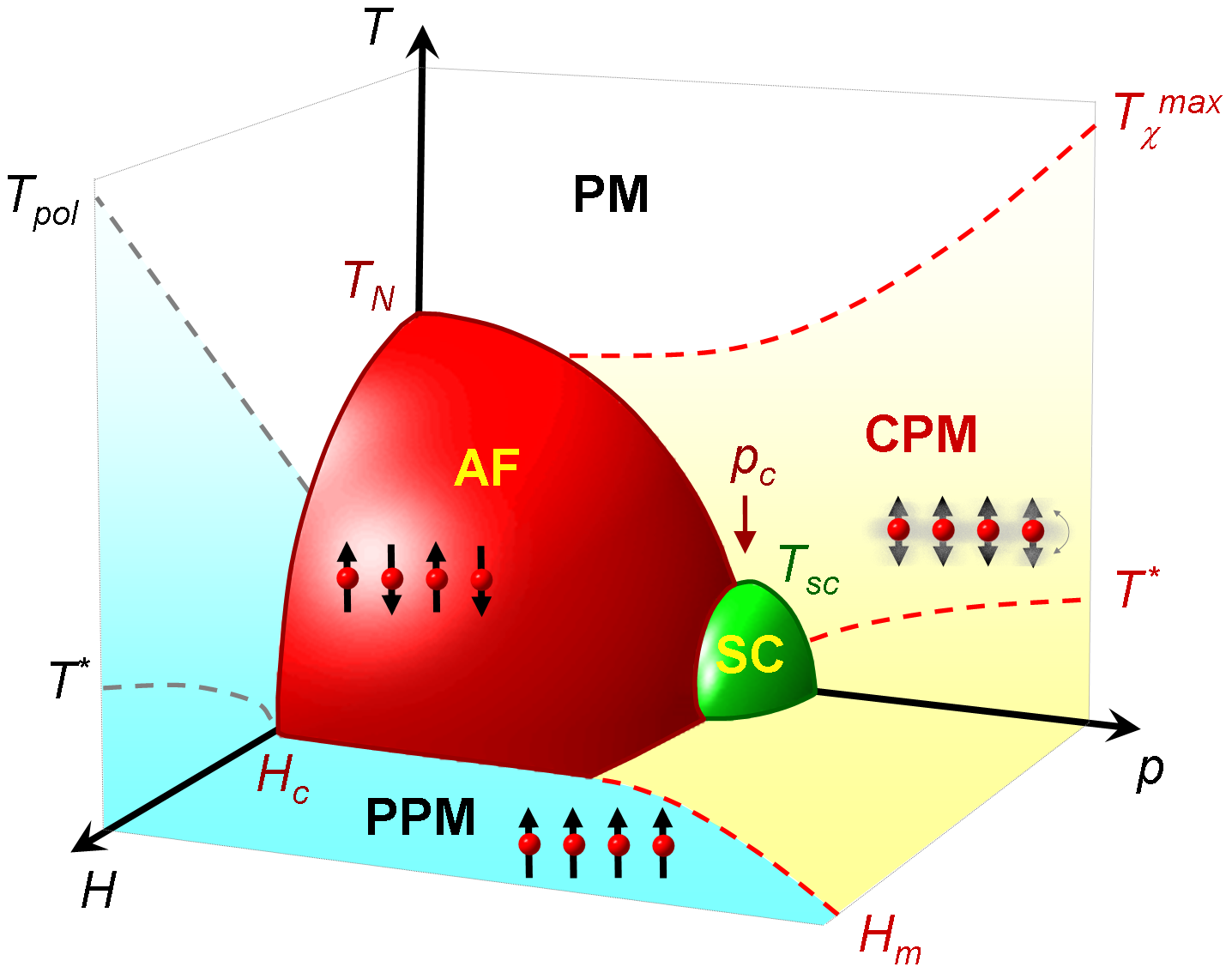}
\caption{Schematic magnetic field - pressure - temperature phase diagram of heavy-fermion antiferromagnets. AF, SC, CPM and PPM denote the antiferromagnetic and superconducting phases, the correlated paramagnetic and polarized paramagnetic regimes, respectively. Sketches of antiferromagnetic moments, polarized moments, and fluctuating moments (assuming Ising magnetic anisotropy) are presented for the AF phase and the PPM and CPM regimes, respectively. All parameters $T_N$, $T_{\chi}^{max}$, $T_{pol}$, $T_{sc}$, $p_c$, $H_c$, and $H_m$ presented in this phase diagram have been defined in Section I.}
\label{Fig1}
\vspace{3 mm}
\end{figure}

The importance of quantum criticality \cite{Hertz76} has been emphasized for a large variety of electronic materials, ranging from high-temperature - cuprate \cite{Valla99} and iron-based \cite{shibauchi14} - superconductors, heavy-fermion systems \cite{knafo09}, to low-dimensional quantum magnets \cite{coldea10,merchant14}, tuned by chemical doping, high pressure, or intense magnetic field. In many of these systems both pressure (or doping) and magnetic field can destabilize a magnetically-ordered phase and lead to critical non-Fermi liquid behaviors \cite{stewart01,cooper09}. However, the critical properties they induce are not necessarily equivalent. While pressure usually drives to a quantum phase transition between a magnetically-ordered state and a paramagnetic (PM) regime, a magnetic field leads to drastically-different effects caused by the polarization of the magnetic moments along the field direction. The comparison between the effects of magnetic field and pressure on quantum criticality has been scarcely explored so far, mainly due to the experimental challenges of such experiments under combined extreme conditions, to which one should add a third dimension, the temperature. For this purpose, heavy-fermion materials are ideal systems: they present the unique advantage of having low electronic energy scales, which allows their quantum critical properties to be tuned by experimentally-accessible pressures and fields. From now on we will focus on Ce-based anisotropic heavy-fermion antiferromagnets, to which belongs CeRh$_2$Si$_2$ investigated here. In the following, we present the main features of their 3D magnetic phase diagram, drawn schematically in Fig. \ref{Fig1}.

Applied to Ce-based antiferromagnets, pressure can tune the electronic correlations and induce a quantum magnetic phase transition at a critical pressure $p_c$, where the antiferromagnetic (AF) ordering temperature $T_N$ collapses to zero and is replaced by a correlated paramagnetic (CPM) regime. Well-defined anomalies (for instance steps or kinks in the magnetic susceptibility $\chi$, the heat capacity $C_p$, and the electrical resistivity $\rho$) are the signature of the phase transitions at $T_N$ and $p_c$. The onset of the CPM regime is a crossover and consists in a progressive change of the physical properties induced by low-temperature electronic correlations. Strong intersite magnetic fluctuations are generally observed in the CPM regime of anisotropic heavy-fermion paramagnets, indicating the proximity of quantum magnetic instabilities \cite{aoki13}. Several characteristic temperatures (instead of a unique critical temperature at a phase transition) can be associated with this crossover to the CPM regime. Its setting in leads to a broad maximum of $\chi$ at the temperature $T_{\chi}^{max}$, below which $\chi$ almost saturates \cite{aoki13,settai07}. At low temperature, the relation $\chi\propto\gamma\propto m^*$, where $\gamma$ is the Sommerfeld coefficient $\gamma=C_p/T$ and $C_p$ is the specific heat, is verified, indicating a Fermi liquid behavior associated with an effective mass $m^*$ \cite{lee86}. The crossover to the CPM regime also leads to a broad maximum at the temperature $T_{\rho}^{max}$ in the electronic (non-phononic) contribution $\rho^{el}$ to the electrical resistivity, which is related with an inflexion point at a temperature $T_{\partial\rho/\partial H}^{max}<T_{\rho}^{max}$ in the resistivity \cite{ohashi03}. A similarity between the broad anomalies observed at $T_{\chi}^{max}$ in $\chi$ and at $T_{\rho}^{max}$ in $\rho^{el}$ has been emphasized \cite{scheerer12}. A $T^2$ Fermi-liquid variation of the resistivity, with a quadratic coefficient $A\propto m^{*2}$, is also generally observed at temperatures below $T^*$ \cite{tsujii2005}. These different temperature scales, with a hierarchy $T^*\ll T_{\partial\rho/\partial H}^{max}< T_{\rho}^{max},T_{\chi}^{max}$ generally observed in heavy-fermion paramagnets (cf. the prototypical heavy-fermion paramagnet CeRu$_2$Si$_2$, where $T^*\simeq500$~mK \cite{daou06} and  $T_{\chi}^{max}\simeq10$~K \cite{fisher91}), are all linked to the crossover into the CPM regime. Fermi-liquid behavior is not restricted to the CPM regime and can be observed inside the antiferromagnetic state. At $p_c$, $T^*$ passes through a minimum or vanishes, whereas the resistivity coefficient $A$ is maximal and may even show a divergence \cite{gegenwart08}. Driven by quantum criticality, i.e., usually by critical magnetic fluctuations, heavy effective masses $m^*$ can reach up to a thousand times that of the free electrons \cite{flouquet05,lohneysen07}. In many systems, low-temperature superconductivity mediated by the critical magnetic fluctuations also develops below a temperature $T_{sc}$ in the vicinity of $p_c$ \cite{pfleiderer09}. Interestingly, in several heavy-fermion antiferromagnets, a broad maximum of susceptibility is observed at a temperature $T_{\chi}^{max}$ higher than the N\'{e}el temperature $T_N$, indicating a CPM regime preceding the AF ordering \cite{aoki13,settai07} (see Fig. \ref{Fig1} at pressures just below $p_c$).

Quantum criticality can also be reached by applying magnetic field, which destroys magnetic order and leads to a quantum phase transition to a polarized paramagnet (PPM) regime, above which a large part of the magnetic moments is aligned, at a critical field $H_c$ \cite{aoki13,friedemann09}. In strongly-anisotropic heavy-fermion antiferromagnets, $H_c$ varies significantly with the field-direction and spin-flop transitions \cite{note_SF} are forbidden. This contrasts with the case of isotropic or nearly-isotropic antiferromagnets, where $H_c$ is almost-independent of the field direction and where spin-flop transitions \cite{grube07} (or crossovers controlled by a spin-flop-like domain alignment \cite{knafo07}) can be induced at low field. When the Ising anisotropy is strong, a magnetic field applied along the easy axis induces a first-order metamagnetic transition at $H_c$. Well-defined anomalies are induced at the transition field $H_c$ in measured physical quantities, such as a jump in the magnetization driven by a sudden polarization of the moments along the field direction \cite{stryjewski05,settai97}, or a sudden step in $\rho$ at low temperature, which is replaced by kink when the temperature is raised \cite{knafo10}. Similarly to the pressure effect, an enhancement of the resistivity coefficient $A$ and a vanishing of $T^*$ have been reported at $H_c$ \cite{gegenwart08}. In anisotropic heavy-fermion paramagnets, a magnetic field applied along the easy axis also leads to a PPM regime, at a pseudo-metamagnetic field usually noted $H_m$ \cite{aoki13,note_Hc_Hm}. $H_m$ corresponds to a magnetic crossover and leads to broad anomalies in measured physical quantities, as a rounded step in $\rho$ at low-temperature, which transforms into a broad maximum when the temperature is raised \cite{daou06}. Although they are both paramagnetic, the low-temperature CPM and PPM regimes may differ significantly. The CPM regime is characterized by strong electronic correlations and a small magnetic polarization under magnetic field, while the PPM regime is associated with a strong magnetic polarization and reduced intersite electronic correlations. This difference is highlighted in the prototypical heavy-fermion paramagnet CeRu$_2$Si$_2$, where antiferromagnetic fluctuations are present in the CPM state but vanish in the PPM regime \cite{rossat88,flouquet04}. Under high fields, a high-temperature scale $T_{pol}$ also indicates the crossover from the low-temperature PPM regime to the high-temperature paramagnetic regime, being associated with a progressive loss of the field-induced magnetic polarization when the temperature is raised.

In the Ising antiferromagnet CeRh$_2$Si$_2$ investigated here, the N\'eel temperature $T_N$ of 36~K at ambient pressure, below which AF moments are aligned along $\mathbf{c}$ \cite{grier84}, vanishes under a pressure $p_c\simeq1$~GPa \cite{kawarasaki00,muramatsu99}. Superconductivity appears in a restricted pressure window around $p_c$ \cite{movshovich96,araki02}. A magnetic field $\mu_0H_c=26$~T applied along $\mathbf{c}$ induces a first-order transition into the PPM regime \cite{settai97,abe97,knafo10}, which is pushed to higher fields under pressure \cite{hamamoto00}. Pressure- and magnetic-field-induced changes of the magnetic structure inside the AF phase have also been reported \cite{kawarasaki00,knafo10,araki02}. Contrary to most other Ce-systems where a non-Fermi-liquid behavior is found, a low-temperature Fermi-liquid $T^2$ dependence of the resistivity is reported in CeRh$_2$Si$_2$ at all pressures and magnetic fields \cite{araki02,knafo10}, including at $p_c$ and $H_c$ where first-order transitions occur. Although the superconducting (SC), PM and/or AF phases boundaries have been determined for several compounds in restricted pressure, field, and temperature windows \cite{aoki13,friedemann09,knebel11,zaum05,matsubayashi15}, a full ($H$,$p$,$T$) phase diagram, including the temperature and pressure evolution of the critical lines $H_c$ and $H_m$, had not been established yet for a pure stoichiometric compound.

Here, we explore the complete three-dimensional ($H$,$p$,$T$) phase diagram of CeRh$_2$Si$_2$, and we extract its magnetic-field- and pressure-induced quantum critical properties. In particular, the need to consider carefully the high-temperature properties to understand the complex quantum critical properties in the ($H$,$p$) plane, i.e., at the borderlines between the AF, CPM, and PPM quantum states, is emphasized. After an introduction to the experimental techniques in Section \ref{sect2}, magnetoresistivity data and the 3D magnetic phase diagram obtained here are presented in Section \ref{sect3}. A comparison of the temperature and magnetic field scales, and an extraction of the quantum critical properties (via the quadratic coefficient $A$ in the resistivity), are presented in Section \ref{sect4}. Finally, discussion and conclusion are made in Sections \ref{sect5} and \ref{sect6}, respectively.

\section{Experimental techniques}
\label{sect2}

The single crystals studied here were grown by the Czochralsky method in a tetra-arc furnace. Magnetoresistivity measurements have been performed under combined extreme conditions of high pressure up to 4~GPa,  pulsed magnetic fields up to 60~T, and temperatures down to 1.4~K. The results presented here correspond to four sets of experiments under pulsed magnetic fields (using four single crystals: samples $\sharp1$, $\sharp2$, $\sharp3$, and $\sharp4$): the first one at ambient pressure (cf. \cite{knafo10} for the experiment details), and the three others using a pressure cell. Residual resistivity ratios $\rho_{x,x}(T=300\mathrm{K})/\rho_{x,x}(T\rightarrow0)$ of $\simeq 50$, 40, 145, and 25 were found for samples $\sharp1$, $\sharp2$, $\sharp3$, and $\sharp4$, respectively, at ambient pressure. Two new Bridgman-type pressure cells specially designed - at the Niigata University (cell "1") and the CEA Grenoble (cell "2") - for the pulsed fields have been used in 20-mm bore $^4$He cryostats under long-duration (rise of 50~ms and fall of 330~ms) pulsed magnetic fields up to 60~T generated at the LNCMI-Toulouse. The electrical resistivity was measured by the four-point technique, at frequencies from 20 to 70~kHz, with a current of 10~mA. Data have been analyzed using digital lock-ins (developed at the LNCMI-T by Hanappel and Fabr\`{e}ges). Cell "1" was used for a set of measurements on sample $\sharp2$, at the pressures $p=0.65$, 0.75, 1, 1.2, 1.5, 2, 2.5 and 3~GPa, and cell "2" was used to study sample $\sharp3$ at the pressures $p=0.95$, 2, 3, and 4~GPa, and sample $\sharp4$ at the pressures $p=0.65$, 0.75, 0.85, and 0.9~GPa. The pressure was estimated by checking the superconducting transition from the resistivity of a lead sample, placed into the cell close to the CeRh$_2$Si$_2$ sample, and by comparing the temperature dependence of $\rho_{x,x}$ measured here with data from \cite{ohashi02,ohashi03}. To perform experiments under pulsed magnetic fields combined with pressures as high as 4~GPa, the price to pay was to accept small heating effects induced on the sample by the metallic body of the cell, whose other parts (anvils, gasket, etc.) are non-metallic, during the pulsed-field shots. The presented data were recorded during the rise of the field pulses, where the sample heating remains acceptable. Assuming a linear increase of the temperature versus time during the pulse, the temperature of the sample has been corrected to extract the $A$ coefficient and the magnetic phase diagram from the resistivity data. Details about the pressure cells, but also the procedure to estimate the heating effects and correct the sample temperature during the field pulses, can be found in Refs. \cite{braithwaite16,settai15}.

\section{Magnetoresistivity and magnetic phase diagram}
\label{sect3}

\begin{figure}[h]
\includegraphics[width=1.0\columnwidth]{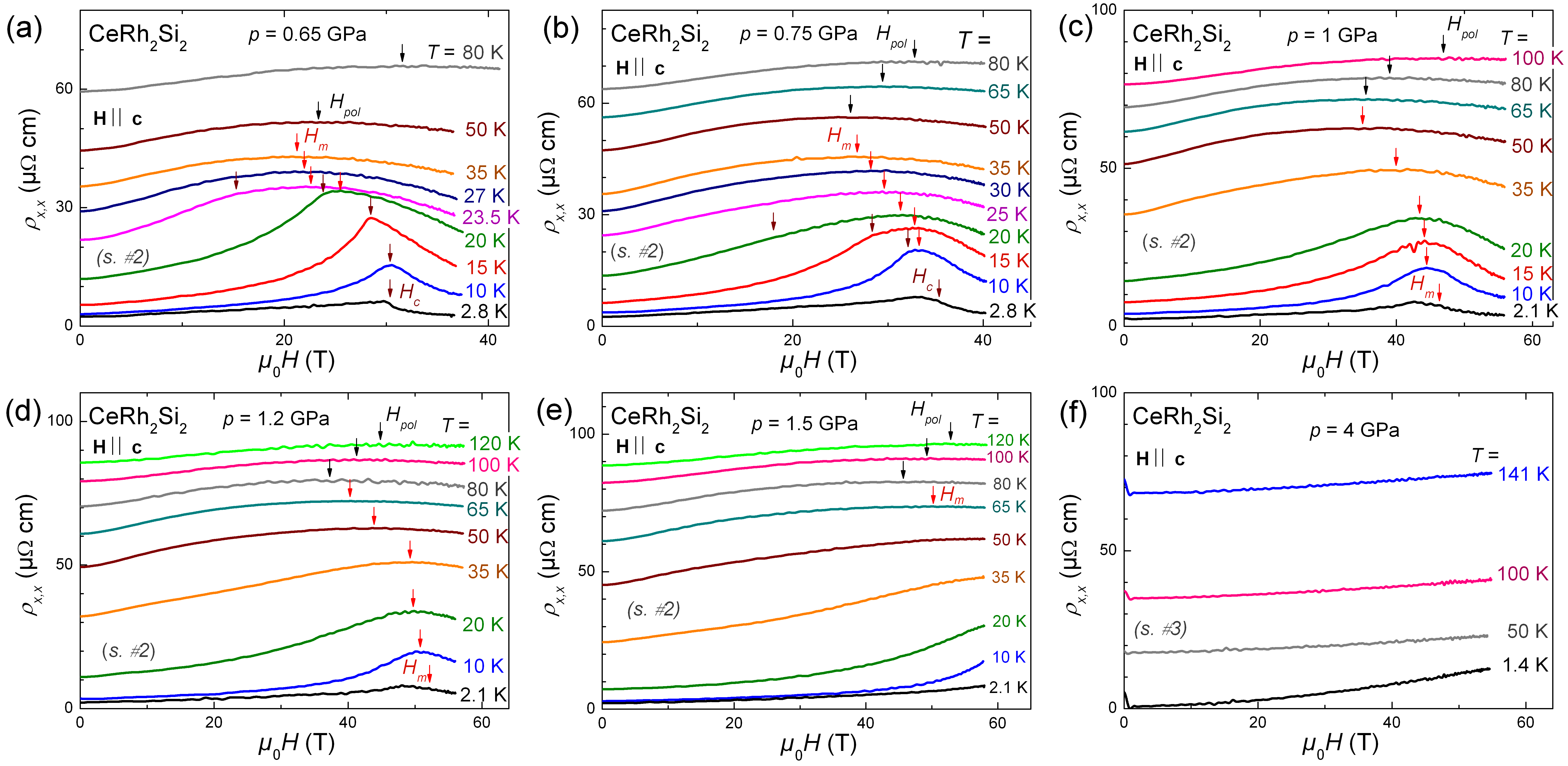}
\caption{$\rho_{x,x}$ of CeRh$_2$Si$_2$ under pressures combined with fields up to 60~T applied along $\mathbf{c}$ and temperatures $1.4\leq T\leq141$~K. Data are presented at (a) $p=0.65$~GPa (sample $\sharp2$), (b) $p=0.75$~GPa (sample $\sharp2$), (c) $p=1$~GPa (sample $\sharp2$), (d) $p=1.2$~GPa (sample $\sharp2$), (e) $p=1.5$~GPa (sample $\sharp2$), and(e) $p=4$~GPa (sample $\sharp3$).}
\label{Fig2}
\end{figure}

\begin{figure}[h]
\includegraphics[width=1\columnwidth]{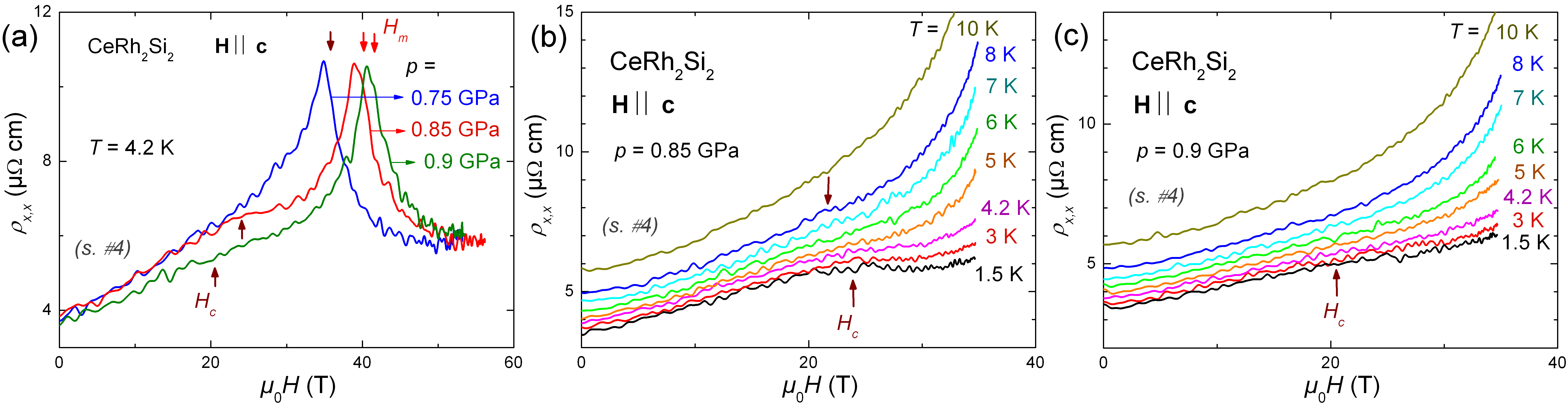}
\caption{Magnetic-field-dependence of the resistivity $\rho_{x,x}$ of CeRh$_2$Si$_2$ emphasizing the low-temperature decoupling of $H_c$ and $H_m$ at pressures just below $p_c$ (sample $\sharp4$). $\rho_{x,x}$ versus $H$ (a) at $T=4.2$~K and $p=0.75$, 0.85, and 0.9~GPa, and at $1.5\leq T\leq10$~K, in fields $\mu_0H\leq35$~T: (b) at $p=0.85$~GPa and (c) at $p=0.9$~GPa.}
\label{Fig3}
\end{figure}

The in-plane magnetoresistivity $\rho_{x,x}$ of CeRh$_2$Si$_2$ under magnetic fields up to 60~T is shown in Fig. \ref{Fig2} for a wide range of pressures and temperatures (cf. \cite{knafo10} for a similar study at ambient pressure). In spite of different residual resistivity ratios varying from 25 to 145, the magnetoresistivity of the four samples investigated here shows similar features, indicating their magnetic - but not orbital - origin (see Ref. \cite{note_orb} and Appendix). For $p<p_c$ (Figs. \ref{Fig2}(a-b), $p=0.65$ and 0.75~GPa) and at low temperature, a step in $\rho_{x,x}$ vs $H$ is seen at the AF borderline $H_c$, which coincides with $H_m$. At intermediate temperatures, a decoupling of $H_c$ (kink) and $H_m$ (broad maximum) is seen for $T>T_x=20$ and 10~K, at $p=0.65$ and 0.75~GPa, respectively. The maximum of $\rho_{x,x}$, which is ascribed to the pseudo-metamagnetic field $H_m$, survives above $T_N$ at pressures $p<p_c\simeq1$~GPa and can be observed down to the lowest temperatures at pressures $p_c\leq p \leq1.2$~GPa (see Figs. \ref{Fig2}(c-d)). At low pressures $p \ll p_c$, the decoupling of $H_c$ and $H_m$ can be observed only in a narrow temperature range $T_x \leq T \leq T_N$, and it is active down to the lowest temperature for pressures just below $p_c$, as shown in Fig. \ref{Fig3} for the pressures $p=0.85$ and 0.9~GPa. Conversely to $T_N$, $\mu_0H_c$ increases under pressure, from 26~T at $p=1$~bar to 35~T at $p=0.75$~GPa. For $p \gtrsim p_c$, $H_m$ at low temperatures is found to increase, reaching 51.5~T at $p=1.2$~GPa (see Fig. \ref{Fig2}(d)). $H_m$ decreases with increasing temperature and we lose its trace at temperatures higher than 50 and 80~K, at the pressures $p=0.75$ and 1.2~GPa, respectively. At high temperature, another broad maximum in $\rho_{x,x}$ is observed at the field $H_{pol}$, which increases with increasing $T$. $H_{pol}$, which is equivalent to $T_{pol}$ in Fig. \ref{Fig1}, is a signature of the crossover between the low-field high-temperature PM and the high-field PPM regimes. At $p=1.5$~GPa and low temperature, $H_m$ is beyond the experimental window [0;60~T], and we can only see its trace, as well as that of $H_{pol}$, at temperatures above 65~K (see Fig. \ref{Fig2}(e)). At $p=4$~GPa, no trace of $H_m$ nor $H_{pol}$ can be found up to 141 K, due to field scales far beyond the accessible experimental window (see Fig. \ref{Fig2}(f)).

\onecolumngrid
\vspace{10 mm}

\twocolumngrid

\begin{figure}[h]
\includegraphics[width=0.72\columnwidth]{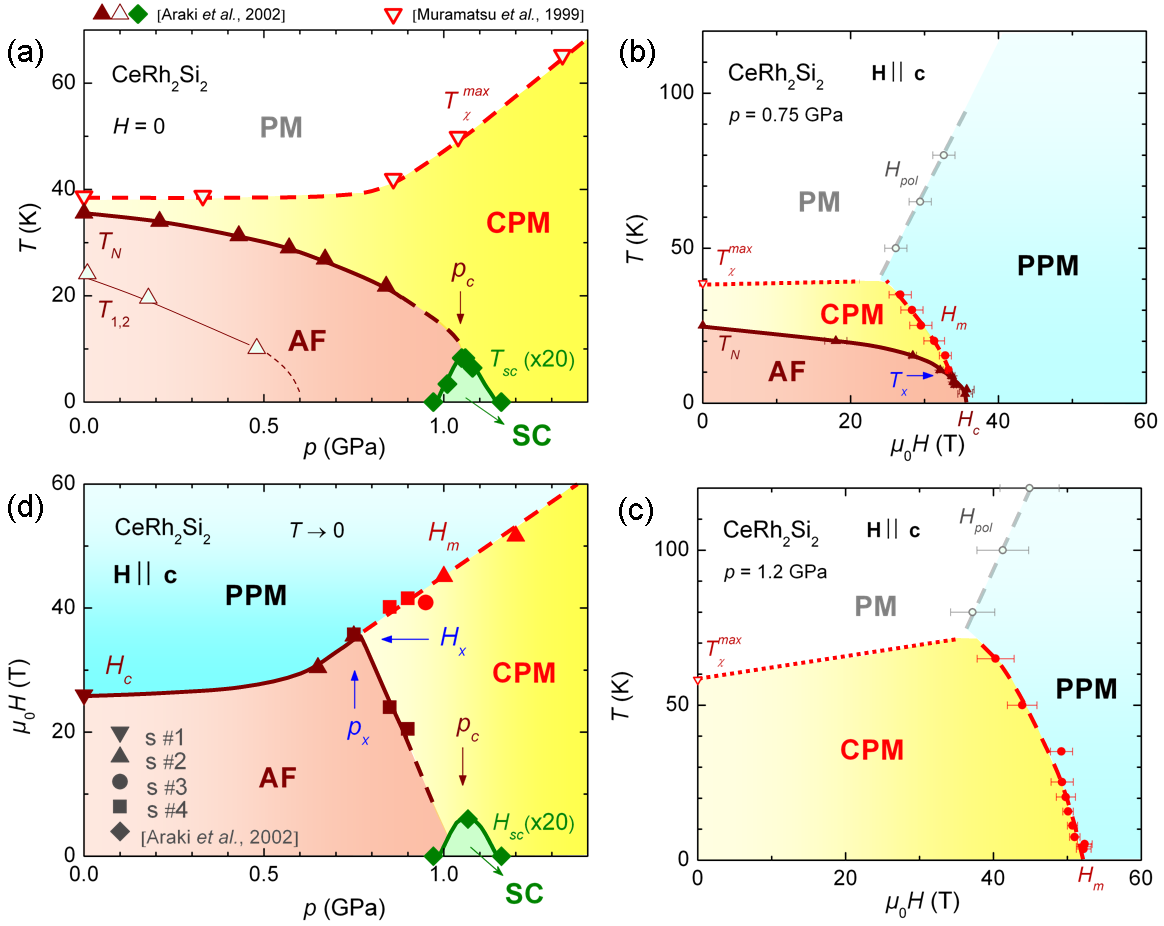}
\caption{($H$,$p$,$T$) phase diagram of CeRh$_2$Si$_2$ under a magnetic field $\mathbf{H}\parallel \mathbf{c}$. (a) ($p$,$T$) phase diagram at $H=0$. (b-c) ($H$,$T$) phase diagrams at $p=0.75$ and 1.2~GPa. (d) ($p$,$H$) phase diagram in the limit $T\rightarrow0$.}
\label{Fig4}
\end{figure}

Fig. \ref{Fig4}(a) presents the zero-field ($p$,$T$) phase diagram of CeRh$_2$Si$_2$ (constructed using susceptibility data from Ref. [\onlinecite{muramatsu99}] and resistivity data from Ref. [\onlinecite{araki02}]), showing that antiferromagnetism vanishes at $p_c\simeq1$~GPa, where superconductivity develops at very low temperature. $T_\chi^{max}$ is almost constant for $p<0.8$~GPa and increases almost linearly with $p$ for $p>0.8$~GPa. Hence the CPM regime, which exists at temperatures $T_N<T\lesssim T_\chi^{max}$, becomes more extended under pressure and is the low-temperature ground state for $p>p_c$. Figs. \ref{Fig4}(b-c) show the ($H$,$T$) phase diagrams obtained here at $p=0.75$ and 1.2~GPa. At $p=0.75$ GPa and at low temperature, antiferromagnetism vanishes at $\mu_0H_c \simeq35$~T, above which the system is polarized paramagnetically. At temperatures $T>T_x=10$~K, $H_c$ and $H_m$ become separated. The temperature of $\simeq40$~K at which $H_m$ vanishes and $H_{pol}$ appears coincides with the temperature scale $T_\chi^{max}$. At $p=0.75$ GPa, the CPM regime is, thus, established at temperatures $T_N<T\lesssim T_\chi^{max}$ and magnetic fields up to $H_m$. At pressures $p\gtrsim p_c$, the ground state is the CPM regime. The temperature at which its boundary $H_m$ falls and $H_{pol}$ appears is also similar to $T_\chi^{max}$. This indicates that the CPM regime can be delimited by the crossover scales $T_\chi^{max}$ and $H_m$, which both increase with increasing pressure. When $T>T_\chi^{max}$, $H_{pol}$ increases linearly with $T$ at all pressures, as observed at $p=1$~bar \cite{knafo10}, and shows little pressure dependence. Fig. \ref{Fig4}(d) presents the ($p$,$H$) phase diagram extracted in the limit $T\rightarrow0$, showing that the AF phase boundary $H_c$ slowly increases with $p$ for $p<p_x=0.75$~GPa, before suddenly decreasing above $p_x$. The borderline of the zero-resistivity SC state $\mu_0H_{SC}$, which reaches $\simeq0.3$~T for $T\rightarrow0$ (data from \cite{araki02}), is also plotted in Fig. \ref{Fig4}(d). The critical point ($p_x$,$H_x$), where $\mu_0H_x=35$~T, separates the AF phase, the high-pressure CPM regime, and the high-field PPM regime in the limit $T\rightarrow0$. Interestingly, the pseudo-metamagnetic field $H_m$ increases linearly with $p$ in a similar manner to $T_\chi^{max}$ for $p>p_x$ (see Figs. \ref{Fig4}(a,d)).

\vspace{4 mm}

\onecolumngrid

\begin{figure}[h]
\includegraphics[width=0.93\columnwidth]{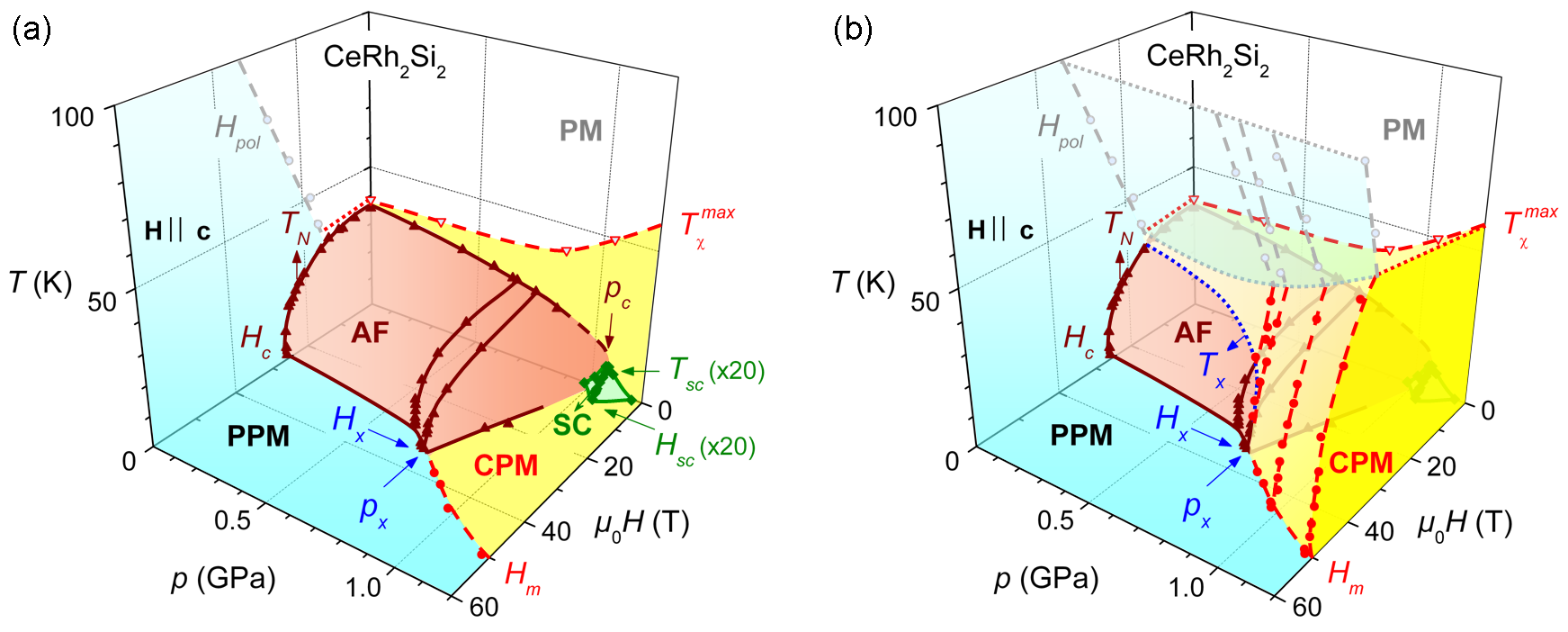}
\caption{3D ($H$,$p$,$T$) phase diagram of CeRh$_2$Si$_2$ under a magnetic field $\mathbf{H}\parallel \mathbf{c}$, for $\mu_0H\leq60$~T, $p\leq1.2$~GPa, and $1.4\leq T\leq100$~K. $T_N$, $T_{1,2}$, and $T_{sc}$ at zero-field and $H_{sc}$ are reproduced from \cite{araki02}, $T_\chi^{max}$ at zero-field is reproduced from \cite{muramatsu99}. (a) emphasizes the 3D views of the AF and SC phases, while (b) emphasizes the 3D views of the CPM and PPM regimes.}
\label{Fig5}
\end{figure}

\twocolumngrid

Fig. \ref{Fig5} shows the three-dimensional magnetic field - pressure - temperature phase diagram of CeRh$_2$Si$_2$ extracted from our magnetoresistivity measurements (but also including data from Refs. \cite{muramatsu99,knafo10,araki02}). Fig. \ref{Fig5}(a) presents a 3D view of the AF and SC phases, while Fig. \ref{Fig5}(b) extends the 3D view to the CPM and PPM regimes. These plots permit to show that the CPM regime englobes the AF phase as long as $T>T_x$ at pressures $p<p_c$. Fig. \ref{Fig5}(b) emphasizes that the temperature $T_x$ separating the three low-temperature phases decreases under pressure and vanishes at the critical point ($H_x$,$p_x$,$T\rightarrow0$). The decoupling of $H_c$ and $H_m$ and its intimate relationship with the decoupling of $T_N$ and $T_\chi^{max}$ are the central features of this phase diagram. We note that the high-temperature border $H_{pol}$ of the PPM regime is almost pressure-independent and defines a high-temperature plane ending on the CPM border.

\section{Energy scales and quantum critical properties}
\label{sect4}

\begin{figure}[t]
\includegraphics[width=0.9\columnwidth]{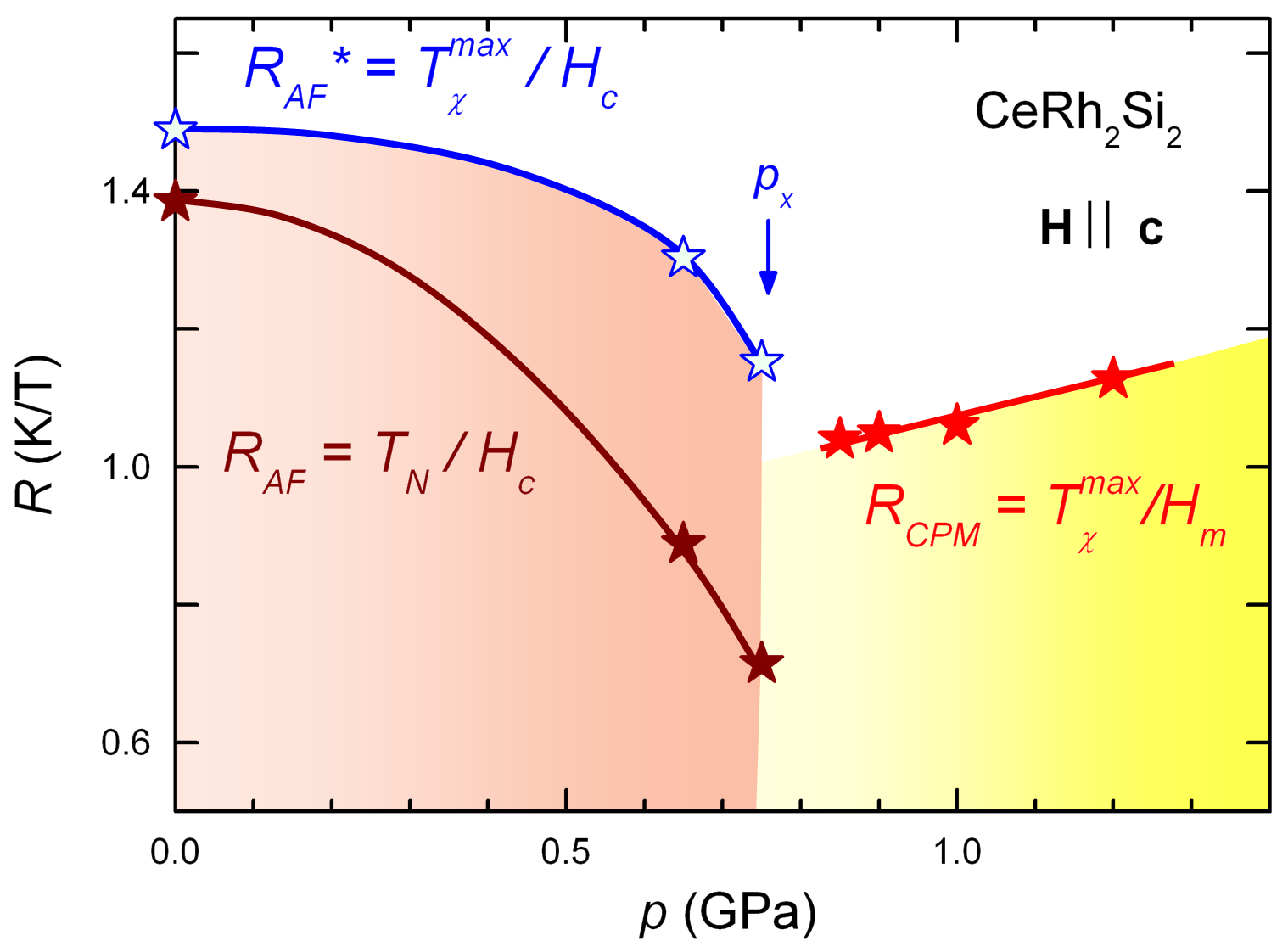}
\caption{Pressure dependence of $R_{AF}=T_N/H_c$ and $R_{AF}^*=T_\chi^{max}/H_c$ for $p<p_x$, and $R_{CPM}=T_\chi^{max}/H_m$ for $p>p_x$.}
\label{Fig6}
\end{figure}

We can better understand the 3D phase diagram by comparing the pressure dependences of the different energy scales. Fig. \ref{Fig6} shows the pressure dependence of the ratios $R_{CPM}=T_\chi^{max}(H=0)/H_m(T\rightarrow0)$ and $R_{AF}=T_N(H=0)/H_c(T\rightarrow0)$ extracted for CeRh$_2$Si$_2$ in its CPM regime and AF phase, respectively. While $R_{CPM}\simeq1.1$~K/T is almost pressure-independent for $p\geq p_x$, $R_{AF}$ strongly varies with $p$ for $p\leq p_x$. This is consistent with a trend already observed for other heavy-fermion systems \cite{aoki13}. The universal constant $R_{CPM}=T_\chi^{max}/H_m\simeq1$~K/T characterizes many heavy-fermion paramagnets \cite{aoki13,onuki04} and indicates that $T_\chi^{max}$ and $H_m$ are controlled by a single parameter. In heavy-fermion paramagnets, the magnetic susceptibility almost saturates at temperatures below $T_\chi^{max}$ and the low-field magnetization $M(H)$ is linear in fields up to almost $H_m$. Pseudo-metamagnetism occurs, thus, at a similar critical value of the magnetization ($M_c\simeq0.5-1$~$\mu_B$) in heavy-fermions paramagnets where $R_{CPM}\simeq1$. Conversely, $R_{AF}=T_N/H_c$ is not a universal constant for heavy-fermion antiferromagnets, where multiple parameters have to be considered \cite{aoki13}. As seen in Figs. \ref{Fig4}(a,d), the increase of $H_c$ with $p$ for $p\leq p_x$ seems to be linked with that of $T_\chi^{max}$, but not with the decrease with $p$ of $T_N$. The ratio $R_{AF}^*=T_\chi^{max}(H=0)/H_c(T\rightarrow0)$, also plotted in Fig. \ref{Fig6} for $p\leq p_x$, is found to vary less than $R_{AF}$, showing that $H_c$ is partly controlled by the magnetic interactions which govern $T_\chi^{max}$.

Having established the phase diagram, we now turn to characterize its quantum critical properties by analyzing the evolution of the effective mass, extracted through the temperature dependence of the resistivity, in the two-dimensional quantum plane ($H$,$p$). At first approximation, we assume that the quadratic coefficient $A$ in the electrical resistivity $\rho_{x,x}=\rho_{x,x}^0+AT^2$, where $\rho_{x,x}^0$ is the residual resistivity, is proportional to $m^{*2}$. Here, a $T^2$ law is observed within the experimental uncertainty in a large temperature window going from 1.5 to $\simeq8-10$~K, for all investigated pressures and magnetic fields. The upper temperature limit of the $T^2$ law is minimal at the critical fields $H_c$ and $H_m$ and pressure $p_c$, where it reaches $\simeq8-10$~K, and is maximal in magnetic fields far from $H_c$ and $H_m$, and pressures far from $p_c$, where it reaches $\simeq15$~K (see Appendix). Fig. \ref{Fig7}(a) presents the field-dependence of $A$ extracted from $\rho_{x,x}$ versus $T$ curves (reconstructed from $\rho_{x,x}$ versus $H$ data measured at constant temperature; see Appendix) at pressures up to 1.5~GPa. Combined magnetic fields and pressures lead to an enhancement of $A$ at $H_c$ (or $H_m$) for $p<p_x$ ($p>p_x$). For $p_x<p<p_c$, an enhancement at $H_c$ is not visible in this $A$ versus $H$ plot because the anomaly at $H_c$ is a rather weak and broad shoulder, in comparison with the anomaly at $H_m$ in the $\rho_{x,x}(H)$ measurement. However, the anomaly at $H_c$ is clearly visible as a well-defined maximum in $A$ versus $p$ plots (see  Fig. \ref{Fig7}(b)) in agreement with the low-temperature ($H$,$p$) phase diagram (Fig. \ref{Fig4}(d)).

Confirming previous reports \cite{knafo10,flouquet10}, the field-driven enhancement of $A$ at ambient pressure is similar to the pressure-driven one at zero-field, the same maximal value of $\simeq25$~$10^{-3}$~$\mu\Omega$cmK$^{-2}$ being reached at ($H_c$, $p=1$~bar) and ($H=0$, $p_c$) (cf.  Fig. \ref{Fig10} in the Appendix). However, quite surprisingly, as pressure is increased, the enhancement of $A$ at the field $H_c$ or $H_m$ becomes considerably larger. This is illustrated in Fig. \ref{Fig7}(c) by the pressure-dependence of $A_{max}$, which is defined as the maximal value of $A$ versus $H$ at a given pressure. Quantum criticality at $H_c$ and $H_m$ leads to the maximal values $A_{max}=A(H_c)$ for $p<p_x$ and $A_{max}=A(H_m)$ for $p>p_x$. We remark that the strong increase of $A_{max}$ with $p$ at pressures $p<p_x$ ends in a maximal value of $A_{max}\simeq150$~$10^{-3}$~$\mu\Omega$cmK$^{-2}$ at ($H_x$,$p_x$), and is followed by an almost pressure-independence of $A_{max}$ for $p\geq p_x$.

\section{Discussion}
\label{sect5}

Here, we present the first study of a pure stoichiometric heavy-fermion antiferromagnet, where the decoupling of two critical fields $H_c$ and $H_m$ is systematically investigated as function of temperature and pressure. Such decoupling of $H_c$ and $H_m$ has also been observed down to $T=100$~mK in the Ising antiferromagnet Ce(Ru$_{0.92}$Rh$_{0.08}$)$_2$Si$_2$ at ambient pressure \cite{aoki12}, and at $T=1.8$~K in Ce$_{0.8}$La$_{0.2}$Ru$_2$Si$_2$ under pressure \cite{haen96}. In the Ising antiferromagnet Ce$_{0.9}$La$_{0.1}$Ru$_2$Si$_2$ at ambient pressure, a decoupling of $H_c$ and $H_m$ has been observed in an intermediate range of temperatures $1.5\leq T \leq T_N=4.4$~K \cite{aoki12,fisher91,haen96}, but not at temperatures below 1.5~K. While pressure and chemical doping generally lead to similar effects on the magnetic phase diagram of heavy-fermion systems, pressure has the advantage to continuously tune the properties of a single crystal without altering its quality. Its combination with intense magnetic fields over a large temperature window permits here to draw the complete ($H$,$p$,$T$) magnetic phase diagram of CeRh$_2$Si$_2$. We show that the low-temperature decoupling of $H_c$ and $H_m$ occurs at pressures just below $p_c$ and is connected with a high-temperature phenomenon, the decoupling of $T_N$ and $T_{\chi}^{max}$. The wrapping of the CPM regime over the AF phase, as seen in the 3D phase diagram in Fig. \ref{Fig5}, is a consequence of the decoupling of these magnetic field and temperature scales. The similarities with studies made on Rh- and La-doped CeRu$_2$Si$_2$ indicate that the complete phase diagram established here for CeRh$_2$Si$_2$ might be generic of a class of heavy-fermion Ising antiferromagnets, where a decoupling of $T_N$ and $T_\chi^{max}$ drives that of $H_c$ and $H_m$.

However, the quantum critical properties at $H_c$ and $H_m$ are sensitive to specific sample properties. In CeRh$_2$Si$_2$, the highest value of $A$ is reached on the $H_m$ critical line, where $A=A(H_m)$ is pressure-independent. Oppositely, in La-doped CeRu$_2$Si$_2$, $A(H_m)$ decreases with $p$ while the Sommerfeld coefficient $\gamma$ ($\propto \sqrt{A}$ in a Fermi liquid picture) is maximal and constant at $H_c$ \cite{aoki11}. Microscopically, we speculate that the enhancement of $A$ at $p_c$ is controlled by critical antiferromagnetic fluctuations (similarly to the Ce$_{1-x}$La$_x$Ru$_2$Si$_2$ case \cite{knafo09}), while that found at $H_m$ may result from additional mechanisms. In CeRu$_2$Si$_2$, critical ferromagnetic fluctuations \cite{aoki13,rossat88,flouquet04} (accompanied by a Fermi surface Lifshitz transition \cite{pourret14}) drive the enhancement of $A$ at $H_m$. Such critical ferromagnetic fluctuations probably play a role in CeRh$_2$Si$_2$ at $H_m$ too. However, in sole magnetic-fluctuations frame there is no obvious reason why the enhancement should be so much larger at $H_m$ than at $H_c$. The relatively 'low' value of the $A$ coefficient extracted here indicates a moderate effective mass in CeRh$_2$Si$_2$. Under pressure, a decrease of the Kadowaki-Woods ratio $A/\gamma^2$ \cite{palacio15} and a reduction of the Ising character of the system \cite{mori99} indicate the progressive recovery of the $N=6$ degeneracy of the $J=5/2$ multiplet, which could result from a Kondo temperature higher than the crystal-field energy scale \cite{hanzawa85}, due to the proximity to a valence transition \cite{watanabe11} (see also \cite{matsuda14}). This picture is supported by the moderate and almost pressure-independent Sommerfeld coefficient $\gamma\simeq80$~mJ/mol/K$^2$ in the high-pressure CPM regime (at zero magnetic field) \cite{graf97}, where a low-temperature Gr\"{u}neisen parameter $\Gamma\simeq7$, i.e., much smaller than in typical heavy-fermion paramagnets (cf. CeRu$_2$Si$_2$ where $\Gamma\simeq200$ \cite{lacerda89}), can also be extracted \cite{note}. An open question is the evolution of the Fermi surface under pressure combined with magnetic field, i.e., through $H_c$ and $H_m$ at pressures close to $p_x$ and $p_c$. At ambient pressure, the Fermi surface established experimentally in the AF phase agrees well with LDA+U band calculations \cite{palacio15,araki01}. On crossing $p_c$, clear marks of Fermi surface change have been established, but a debate remains on the validity of its description by LDA, LDA+U, or some more sophisticated band calculation. Unfortunately, the Fermi surface of the PPM regime in fields above $H_c$ has only be detected by the emergence of a single frequency \cite{sheikinxx}. New attempts to study the Fermi surface, but also valence, under combined pressure and magnetic field, allowing to access the PPM regime in magnetic fields higher than $H_c$ and $H_m$, are now requested.

\onecolumngrid

\begin{figure}[h]
\includegraphics[width=1\columnwidth]{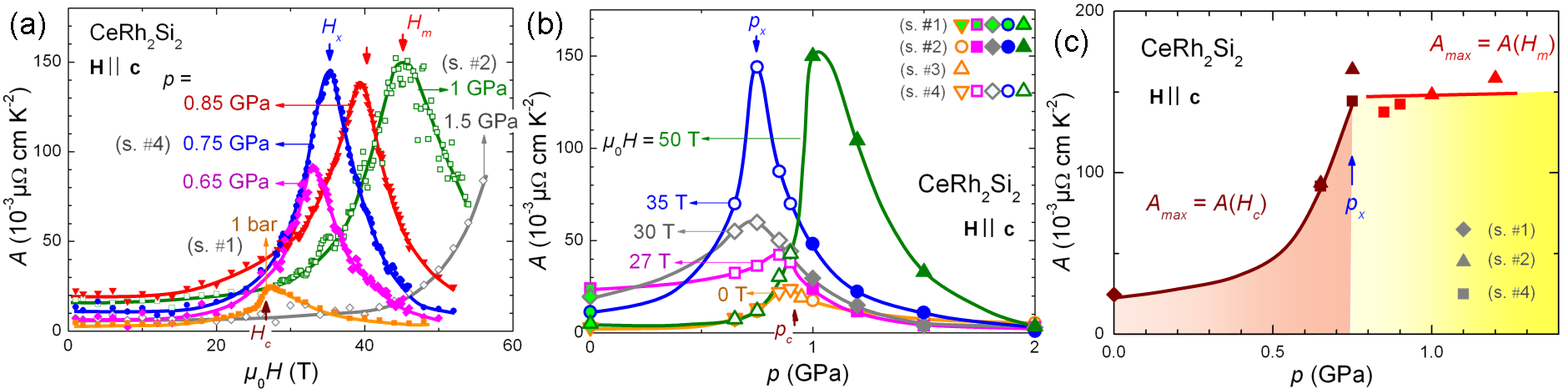}
\caption{Evolution of the Fermi liquid coefficient $A$ as function of magnetic field and pressure. (a) $A$ versus $H$, in magnetic fields up to 55~T, at different pressures. (b) $A$ versus $p$, in pressures up to 2GPa, at different magnetic fields. (c) Pressure-dependence of $A_{max}$ ($=A(H_c)$ for $p<p_x$ and $A(H_m)$ for $p>p_x$.}
\label{Fig7}
\end{figure}

\twocolumngrid

\section{Conclusion}
\label{sect6}

The 3D phase diagram of CeRh$_2$Si$_2$, as well as the pressure- and magnetic-field-variations of its effective mass, confirm the importance of studying over a large temperature window the effects of pressure and magnetic field on quantum critical phenomena. The need to consider the high-temperature properties, such as the onset of antiferromagnetism and correlated paramagnetism, and their intimate relationship with the quantum properties, has been highlighted.

Thanks to a systematic tuning of a 3D phase diagram under combined pressure and magnetic field, we have determined the temperature and pressure dependences of the critical and pseudo-metamagnetic fields $H_c$ and $H_m$, respectively. In particular, our study permitted to show:
\begin{enumerate}
  \item that a low-temperature decoupling of $H_c$ and $H_m$ occurs in a narrow pressure range in the vicinity of the critical pressure $p_c$ (at pressures $p_x<p<p_c$),
  \item that this low-temperature decoupling of $H_c$ and $H_m$ results from a strong high-temperature decoupling of $T_N$ and $T_\chi^{max}$,
  \item that the decoupling of $H_c$ and $H_m$ can also occur in a limited temperature range when $T_N$ and $T_\chi^{max}$ are not sufficiently decoupled (at pressures $p<p_x$).
\end{enumerate}
A significant enhancement of the Fermi liquid coefficient $A$ has also been observed at the critical point ($p_x=0.75$~GPa, $\mu_0H_x=35$~T), which delimits the three low-temperature states (AF phase, CPM and PPM regimes), and on the critical line $H_m$, which separates the CPM and PPM regimes. From a sole magnetic-fluctuations picture, it is difficult to understand why the enhancement of $A$ is much larger at $H_m$ than at $H_c$. Additional effects, like changes of the Fermi surface, valence, and/or magnetic anisotropy, might also play a role in the critical enhancements of $A$. To go beyond the present study and understand the evolution of quantum criticality in the ($p$,$H$) phase diagram, further experiments (e.g., on magnetic fluctuations, valence, Fermi surface etc.) on CeRh$_2$Si$_2$, as well as new theoretical descriptions, are needed.

More generally, the combination of extreme conditions of intense magnetic field and high pressure constitutes a goldmine of novel phenomena in a large variety of strongly-correlated-electron systems. As in heavy-fermion systems, a decoupling of the long-range magnetic ordering and the maximum of the magnetic susceptibility is also observed in low-dimensional quantum magnets \cite{dejongh90} and in high-temperature cuprate superconductors \cite{johnston89,pines13}. Studying the evolution of this decoupling under combined pressure and magnetic field will shed new light on the quantum critical properties of these systems. In the region of optimal doping of cuprate superconductors, where several electronic energy scales fall drastically \cite{pines13}, it will be pertinent to test whether, and if so how, the quantum critical properties could be affected by intense magnetic fields.

\section*{Acknowledgements}

We acknowledge J. B\'{e}ard and M. Nardone for useful discussions. This work has been partly funded by the French ANR contract PRINCESS and the CEFIPRA project 4906 ExtremeSpinLadder.\\

\appendix

\section*{APPENDIX}

\begin{figure}
\includegraphics[width=.67\columnwidth]{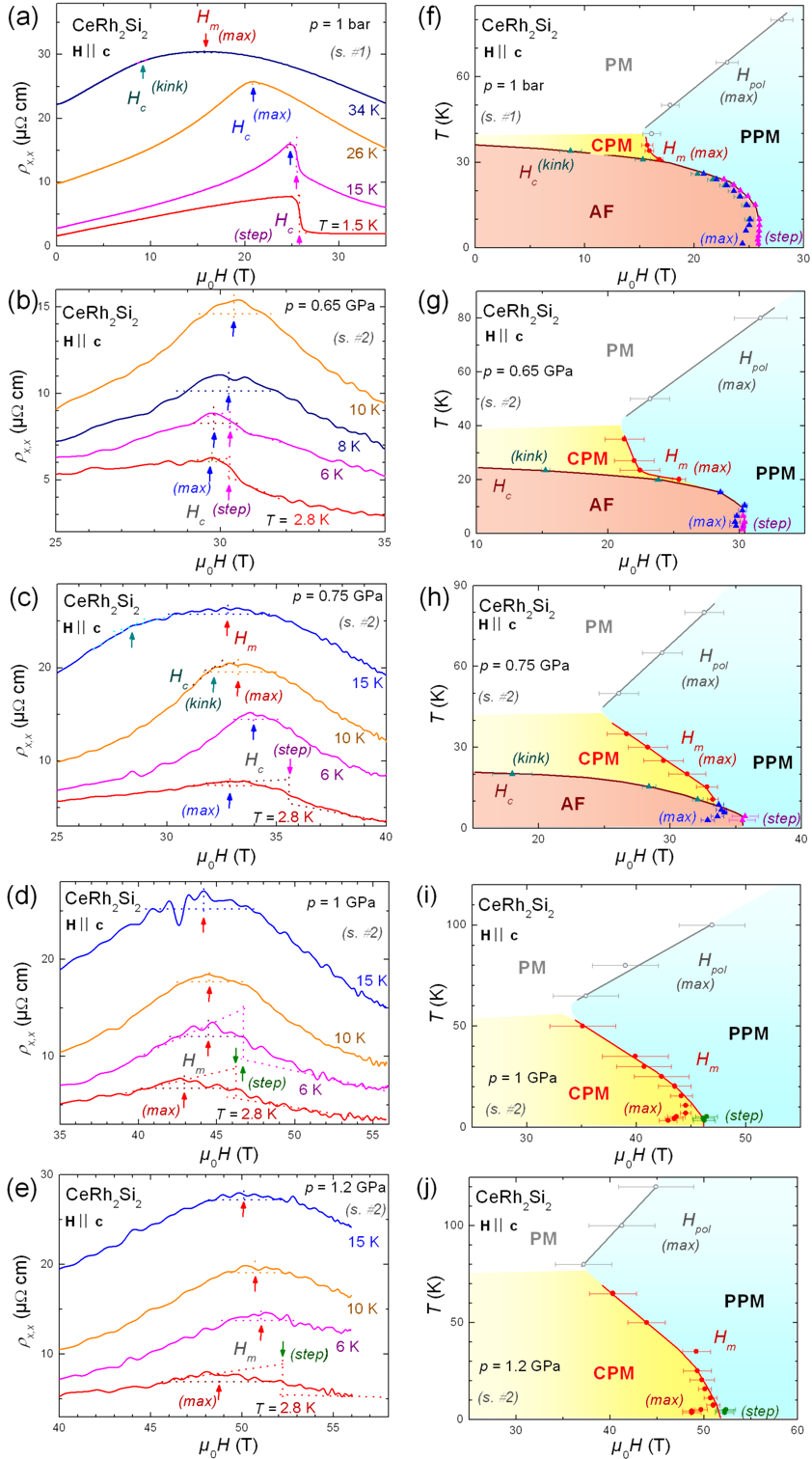}
\caption{Definition of the transitions and crossovers in the magnetoresistivity and phase diagrams extracted at different pressures. (a-e) Magnetoresistivity of sample $\sharp1$ at ambient pressure, and of sample $\sharp2$ at pressures from 0.65 to 1.2~GPa, under pulsed magnetic fields up to 56~T, at various temperatures from 1.5 to 34~K, and (f-j). corresponding magnetic-field - temperature phase diagrams obtained at constant pressures.}
\label{Fig8}
\end{figure}

Fig. \ref{Fig8} presents how the different critical and crossover fields have been defined here. Fig. \ref{Fig8} (a) shows that, at ambient pressure (data from \cite{knafo10}), $H_c$ is characterized by a step at low temperature, which progressively transforms into a maximum and then a kink at higher temperature, in the magnetoresistivity data. At temperatures above $T^*\simeq30$~K and below $T_N(H=0)=36$~K, a broad maximum at a field ascribed to the pseudo-metamagnetic field $H_m$ is decoupled from the kink at the critical field $H_c$. The different anomalies observed the resistivity at ambient pressure are summarized in the magnetic-field - temperature phase diagram in Fig. \ref{Fig8} (f). Fig. \ref{Fig8}(b-e) show the magnetoresistivity obtained here on sample $\sharp2$ at pressures from 0.65 to 1.2~GPa. At low temperature, the anomalies at $H_c$ and $H_m$ under pressure are broadened, in comparison with that at $H_c$ at ambient pressure, presumably because of pressure inhomogeneities in the cells. In the light of other heavy-fermion paramagnets studies (see Ref. \cite{daou06}) and for continuity in the data analysis, $H_m$ has been defined at low temperature at the mid of a step in the resistivity, and at high temperature at the maximum of the resistivity. The magnetic-field - temperature phase diagrams obtained under pressure are plotted in Fig. \ref{Fig8} (g-j). They indicate that the correlated paramagnetic regime is progressively stabilized under pressure, extending from a narrow high-temperature region at ambient pressure to a large region down to the lowest temperature under pressures higher than $p_c\simeq1$~GPa.

\begin{figure}
\includegraphics[width=.7\columnwidth]{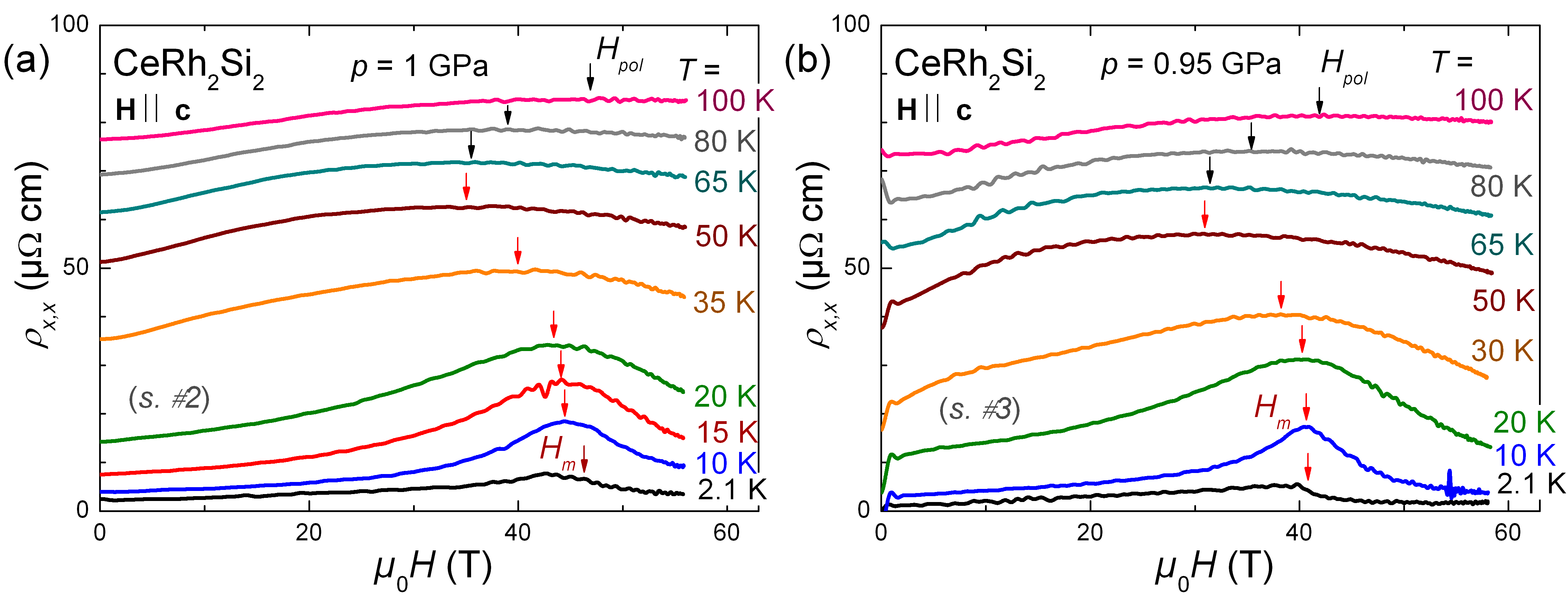}
\caption{Magnetoresistivity (a) of sample $\sharp2$ at $p=1$~GPa and (b) of sample $\sharp3$ at $p=0.95$~GPa at various temperatures from 2.1 to 100~K.}
\label{Fig8b}
\end{figure}

Figs. \ref{Fig8b} (a-b) show the magnetoresistivity versus magnetic field data obtained at similar pressures of 1 and 0.95~GPa on samples $\sharp2$ and $\sharp3$, respectively, for a large set of temperatures from 2.1 to 100~K. This comparison emphasizes the very similar anomalies at $H_m$ and $H_{pol}$ observed in the magnetoresistivity of two samples of very different residual resistivity ratios (of $\simeq40$ for sample $\sharp2$ and $\simeq145$ for sample $\sharp3$), indicating that these anomalies are not controlled by an orbital effect, i.e., the field-induced cyclotron motion of the conduction electrons, and result from the magnetic properties of the system.

\begin{figure}
\includegraphics[width=1\columnwidth]{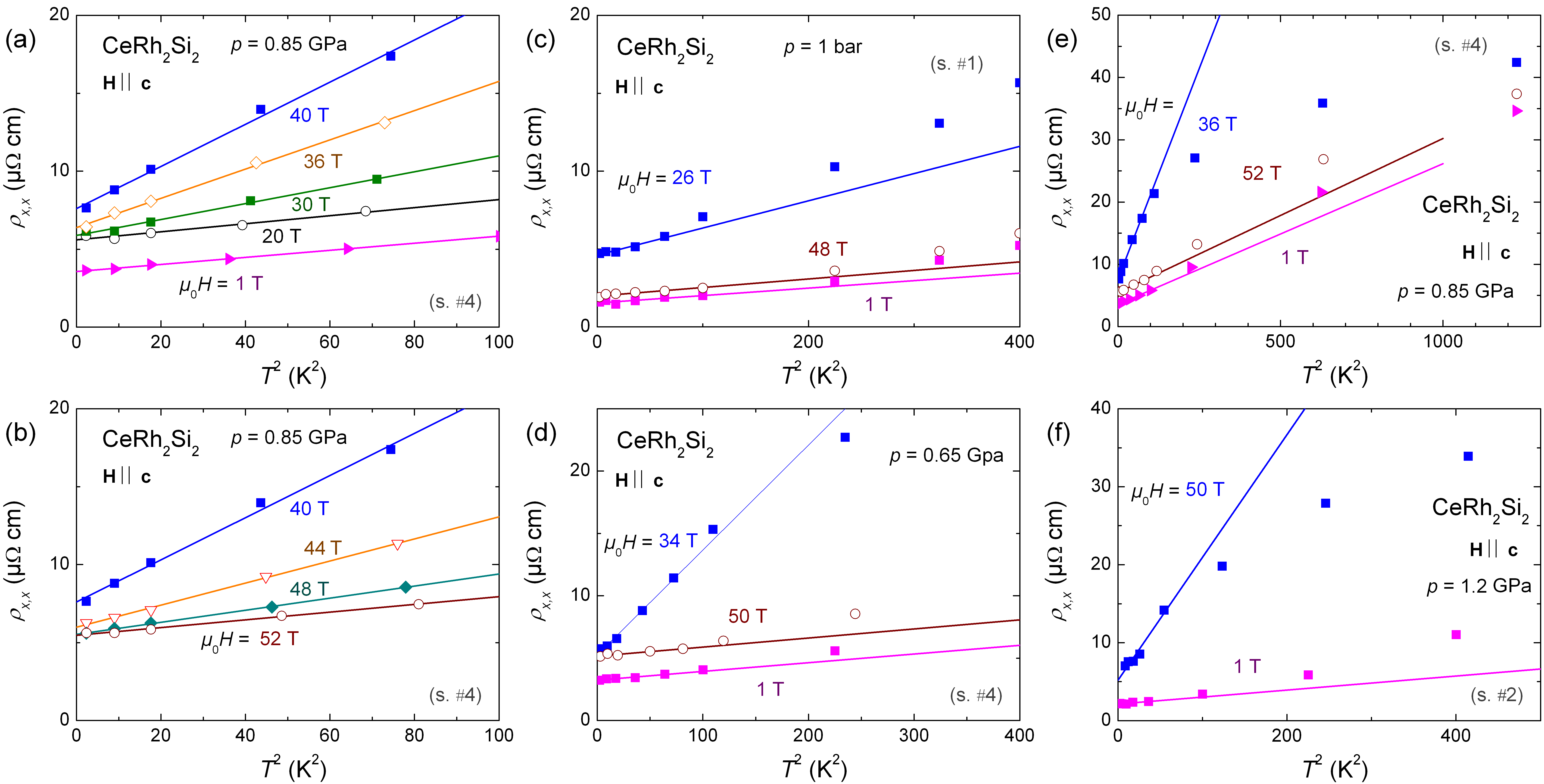}
\caption{(a-b) Resistivity versus $T^2$ of sample $\sharp4$ at $p=0.9$~GPa, for different fields up to 52~T. Resistivity versus $T^2$, in extended temperature scales, of (c) sample $\sharp1$ at $p=1$~bar, (d-e) sample $\sharp4$ at $p=0.65$ and 0.85~GPa, and (f) $\sharp2$ at $p=1.2$~GPa, for selections of magnetic field values from 1 to 52~T. The lines correspond to the fit to the data by a $T^2$ law.}
\label{Fig9}
\end{figure}

For all data, the fit by a $T^2$ law of the magnetoresistivity has been made in a temperature window from 1.5 to 10~K, the temperature having been corrected as described in \cite{braithwaite16}. Fig. \ref{Fig9} (a-b) shows the $T^2$ variation of the resistivity extracted at $p=0.9$~GPa on sample $\sharp4$ using cell "2", for different fields up to 52~T, which indicates a significant enhancement of the quadratic coefficient $A$ at 40~T. Fig. \ref{Fig9} (c-f) shows, in an extended temperature scale, the $T^2$ variation of the resistivity extracted for different values of the magnetic field, at $p=1$~bar, 0.65~GPa, 0.85~GPa, and 1.2 GPa, indicating the departure from the low-temperature $T^2$ behavior at a temperature $T^*$ estimated to $\simeq8-10$~K at $H_c$ and $H_m$, and to $\simeq15$~K at magnetic fields far from $H_c$ and $H_m$.

\begin{figure}
\includegraphics[width=.34\columnwidth]{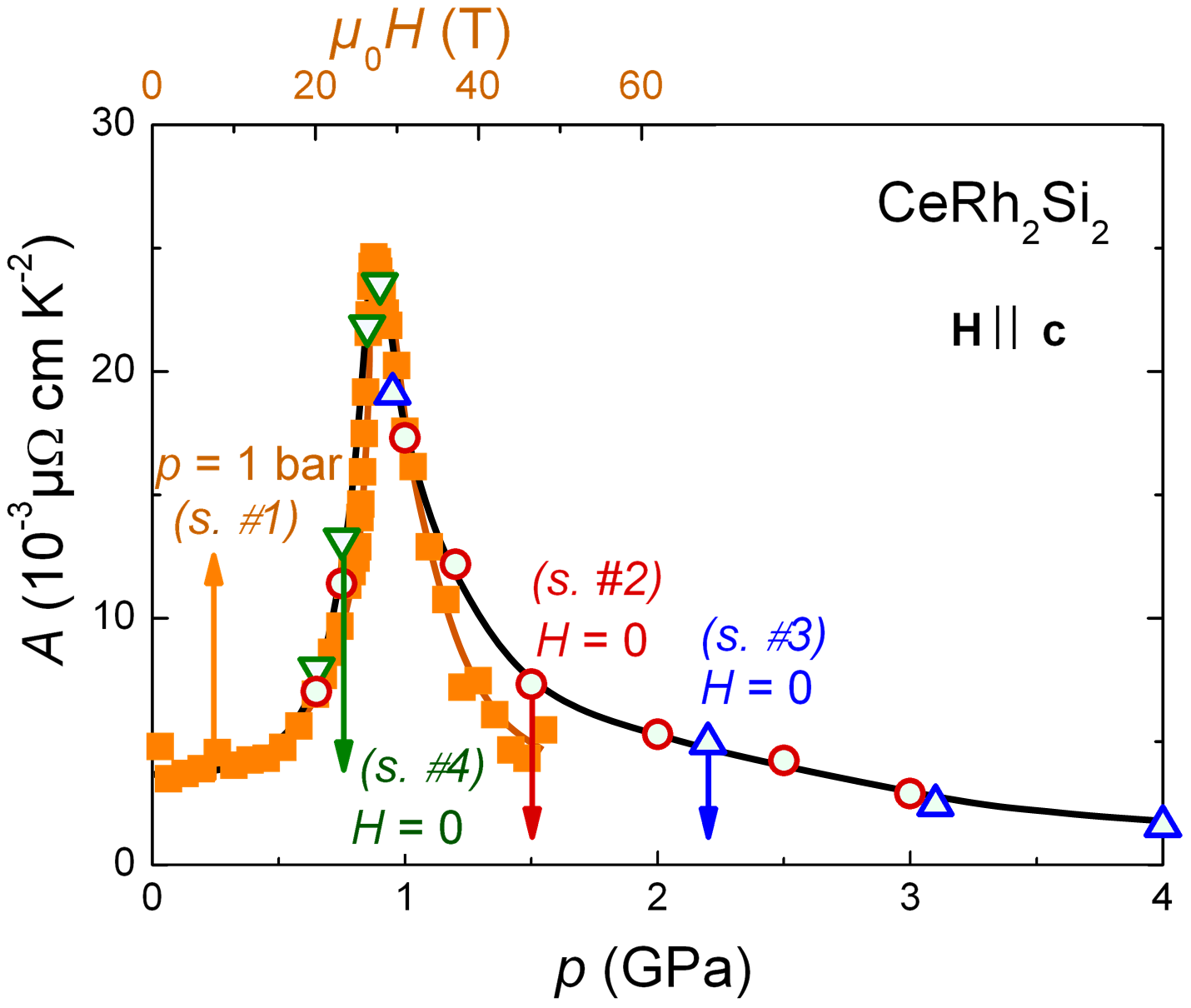}
\caption{Magnetic-field- and pressure-dependence of the quadratic coefficient $A$ of the magnetoresistivity $\rho_{x,x}$ of CeRh$_2$Si$_2$ in $\mathbf{H}\parallel\mathbf{c}$. (a) Comparison of $A$ versus $H$ at $p=1$~bar (sample $\sharp1$) and $A$ versus $p$ at $H=0$ (samples $\sharp2$, $\sharp3$, $\sharp4$).}
\label{Fig10}
\end{figure}

Fig. \ref{Fig10} (a) shows a comparison of the field-dependence (at ambient pressure) and the pressure-dependence (at zero-field) of the coefficient $A$ extracted here for samples $\sharp1$, $\sharp2$, $\sharp3$, and $\sharp4$, confirming the similar enhancements of $A$ up to $\simeq25$~$10^{-3}~\mu\Omega$cm K$^{-2}$ observed at ($H_c$, $p=1$bar) and ($H=0$, $p_c$) reported previously \cite{knafo10,flouquet10}.

\onecolumngrid

\twocolumngrid


\begin{thebibliography}{64}%
\makeatletter
\providecommand \@ifxundefined [1]{%
 \@ifx{#1\undefined}
}%
\providecommand \@ifnum [1]{%
 \ifnum #1\expandafter \@firstoftwo
 \else \expandafter \@secondoftwo
 \fi
}%
\providecommand \@ifx [1]{%
 \ifx #1\expandafter \@firstoftwo
 \else \expandafter \@secondoftwo
 \fi
}%
\providecommand \natexlab [1]{#1}%
\providecommand \enquote  [1]{``#1''}%
\providecommand \bibnamefont  [1]{#1}%
\providecommand \bibfnamefont [1]{#1}%
\providecommand \citenamefont [1]{#1}%
\providecommand \href@noop [0]{\@secondoftwo}%
\providecommand \href [0]{\begingroup \@sanitize@url \@href}%
\providecommand \@href[1]{\@@startlink{#1}\@@href}%
\providecommand \@@href[1]{\endgroup#1\@@endlink}%
\providecommand \@sanitize@url [0]{\catcode `\\12\catcode `\$12\catcode
  `\&12\catcode `\#12\catcode `\^12\catcode `\_12\catcode `\%12\relax}%
\providecommand \@@startlink[1]{}%
\providecommand \@@endlink[0]{}%
\providecommand \url  [0]{\begingroup\@sanitize@url \@url }%
\providecommand \@url [1]{\endgroup\@href {#1}{\urlprefix }}%
\providecommand \urlprefix  [0]{URL }%
\providecommand \Eprint [0]{\href }%
\providecommand \doibase [0]{http://dx.doi.org/}%
\providecommand \selectlanguage [0]{\@gobble}%
\providecommand \bibinfo  [0]{\@secondoftwo}%
\providecommand \bibfield  [0]{\@secondoftwo}%
\providecommand \translation [1]{[#1]}%
\providecommand \BibitemOpen [0]{}%
\providecommand \bibitemStop [0]{}%
\providecommand \bibitemNoStop [0]{.\EOS\space}%
\providecommand \EOS [0]{\spacefactor3000\relax}%
\providecommand \BibitemShut  [1]{\csname bibitem#1\endcsname}%
\let\auto@bib@innerbib\@empty
\bibitem [{\citenamefont {Hertz}(1976)}]{Hertz76}%
  \BibitemOpen
  \bibfield  {author} {\bibinfo {author} {\bibfnamefont {J.~A.}\ \bibnamefont
  {Hertz}},\ }\href {\doibase 10.1103/PhysRevB.14.1165} {\bibfield  {journal}
  {\bibinfo  {journal} {Phys. Rev. B}\ }\textbf {\bibinfo {volume} {14}},\
  \bibinfo {pages} {1165} (\bibinfo {year} {1976})}\BibitemShut {NoStop}%
\bibitem [{\citenamefont {Valla}\ \emph {et~al.}(1999)\citenamefont {Valla},
  \citenamefont {Fedorov}, \citenamefont {Johnson}, \citenamefont {Wells},
  \citenamefont {Hulbert}, \citenamefont {Li}, \citenamefont {Gu},\ and\
  \citenamefont {Koshizuka}}]{Valla99}%
  \BibitemOpen
  \bibfield  {author} {\bibinfo {author} {\bibfnamefont {T.}~\bibnamefont
  {Valla}}, \bibinfo {author} {\bibfnamefont {A.~V.}\ \bibnamefont {Fedorov}},
  \bibinfo {author} {\bibfnamefont {P.~D.}\ \bibnamefont {Johnson}}, \bibinfo
  {author} {\bibfnamefont {B.~O.}\ \bibnamefont {Wells}}, \bibinfo {author}
  {\bibfnamefont {S.~L.}\ \bibnamefont {Hulbert}}, \bibinfo {author}
  {\bibfnamefont {Q.}~\bibnamefont {Li}}, \bibinfo {author} {\bibfnamefont
  {G.~D.}\ \bibnamefont {Gu}}, \ and\ \bibinfo {author} {\bibfnamefont
  {N.}~\bibnamefont {Koshizuka}},\ }\href {\doibase
  10.1126/science.285.5436.2110} {\bibfield  {journal} {\bibinfo  {journal}
  {Science}\ }\textbf {\bibinfo {volume} {285}},\ \bibinfo {pages} {2110}
  (\bibinfo {year} {1999})}\BibitemShut {NoStop}%
\bibitem [{\citenamefont {Shibauchi}\ \emph {et~al.}(2014)\citenamefont
  {Shibauchi}, \citenamefont {Carrington},\ and\ \citenamefont
  {Matsuda}}]{shibauchi14}%
  \BibitemOpen
  \bibfield  {author} {\bibinfo {author} {\bibfnamefont {T.}~\bibnamefont
  {Shibauchi}}, \bibinfo {author} {\bibfnamefont {A.}~\bibnamefont
  {Carrington}}, \ and\ \bibinfo {author} {\bibfnamefont {Y.}~\bibnamefont
  {Matsuda}},\ }\href {\doibase 10.1146/annurev-conmatphys-031113-133921}
  {\bibfield  {journal} {\bibinfo  {journal} {Annu. Rev. Condens. Matter
  Phys.}\ }\textbf {\bibinfo {volume} {5}},\ \bibinfo {pages} {113} (\bibinfo
  {year} {2014})}\BibitemShut {NoStop}%
\bibitem [{\citenamefont {Knafo}\ \emph {et~al.}(2009)\citenamefont {Knafo},
  \citenamefont {Raymond}, \citenamefont {Lejay},\ and\ \citenamefont
  {Flouquet}}]{knafo09}%
  \BibitemOpen
  \bibfield  {author} {\bibinfo {author} {\bibfnamefont {W.}~\bibnamefont
  {Knafo}}, \bibinfo {author} {\bibfnamefont {S.}~\bibnamefont {Raymond}},
  \bibinfo {author} {\bibfnamefont {P.}~\bibnamefont {Lejay}}, \ and\ \bibinfo
  {author} {\bibfnamefont {J.}~\bibnamefont {Flouquet}},\ }\href {\doibase
  10.1038/nphys1374} {\bibfield  {journal} {\bibinfo  {journal} {Nat. Phys.}\
  }\textbf {\bibinfo {volume} {5}},\ \bibinfo {pages} {753} (\bibinfo {year}
  {2009})}\BibitemShut {NoStop}%
\bibitem [{\citenamefont {Coldea}\ \emph {et~al.}(2010)\citenamefont {Coldea},
  \citenamefont {Tennant}, \citenamefont {Wheeler}, \citenamefont {Wawrzynska},
  \citenamefont {Prabhakaran}, \citenamefont {Telling}, \citenamefont
  {Habicht}, \citenamefont {Smeibidl},\ and\ \citenamefont
  {Kiefer}}]{coldea10}%
  \BibitemOpen
  \bibfield  {author} {\bibinfo {author} {\bibfnamefont {R.}~\bibnamefont
  {Coldea}}, \bibinfo {author} {\bibfnamefont {D.~A.}\ \bibnamefont {Tennant}},
  \bibinfo {author} {\bibfnamefont {E.~M.}\ \bibnamefont {Wheeler}}, \bibinfo
  {author} {\bibfnamefont {E.}~\bibnamefont {Wawrzynska}}, \bibinfo {author}
  {\bibfnamefont {D.}~\bibnamefont {Prabhakaran}}, \bibinfo {author}
  {\bibfnamefont {M.}~\bibnamefont {Telling}}, \bibinfo {author} {\bibfnamefont
  {K.}~\bibnamefont {Habicht}}, \bibinfo {author} {\bibfnamefont
  {P.}~\bibnamefont {Smeibidl}}, \ and\ \bibinfo {author} {\bibfnamefont
  {K.}~\bibnamefont {Kiefer}},\ }\href {\doibase 10.1126/science.1180085}
  {\bibfield  {journal} {\bibinfo  {journal} {Science}\ }\textbf {\bibinfo
  {volume} {327}},\ \bibinfo {pages} {177} (\bibinfo {year}
  {2010})}\BibitemShut {NoStop}%
\bibitem [{\citenamefont {Merchant}\ \emph {et~al.}(2014)\citenamefont
  {Merchant}, \citenamefont {Normand}, \citenamefont {Kraemer}, \citenamefont
  {Boehm}, \citenamefont {McMorrow},\ and\ \citenamefont
  {Rueegg}}]{merchant14}%
  \BibitemOpen
  \bibfield  {author} {\bibinfo {author} {\bibfnamefont {P.}~\bibnamefont
  {Merchant}}, \bibinfo {author} {\bibfnamefont {B.}~\bibnamefont {Normand}},
  \bibinfo {author} {\bibfnamefont {K.~W.}\ \bibnamefont {Kraemer}}, \bibinfo
  {author} {\bibfnamefont {M.}~\bibnamefont {Boehm}}, \bibinfo {author}
  {\bibfnamefont {D.~F.}\ \bibnamefont {McMorrow}}, \ and\ \bibinfo {author}
  {\bibfnamefont {C.}~\bibnamefont {Rueegg}},\ }\href@noop {} {\bibfield
  {journal} {\bibinfo  {journal} {Nat. Phys.}\ }\textbf {\bibinfo {volume}
  {10}},\ \bibinfo {pages} {373} (\bibinfo {year} {2014})}\BibitemShut
  {NoStop}%
\bibitem [{\citenamefont {Stewart}(2001)}]{stewart01}%
  \BibitemOpen
  \bibfield  {author} {\bibinfo {author} {\bibfnamefont {G.~R.}\ \bibnamefont
  {Stewart}},\ }\href {\doibase 10.1103/RevModPhys.73.797} {\bibfield
  {journal} {\bibinfo  {journal} {Rev. Mod. Phys.}\ }\textbf {\bibinfo {volume}
  {73}},\ \bibinfo {pages} {797} (\bibinfo {year} {2001})}\BibitemShut
  {NoStop}%
\bibitem [{\citenamefont {Cooper}\ \emph {et~al.}(2009)\citenamefont {Cooper},
  \citenamefont {Wang}, \citenamefont {Vignolle}, \citenamefont {Lipscombe},
  \citenamefont {Hayden}, \citenamefont {Tanabe}, \citenamefont {Adachi},
  \citenamefont {Koike}, \citenamefont {Nohara}, \citenamefont {Takagi},
  \citenamefont {Proust},\ and\ \citenamefont {Hussey}}]{cooper09}%
  \BibitemOpen
  \bibfield  {author} {\bibinfo {author} {\bibfnamefont {R.~A.}\ \bibnamefont
  {Cooper}}, \bibinfo {author} {\bibfnamefont {Y.}~\bibnamefont {Wang}},
  \bibinfo {author} {\bibfnamefont {B.}~\bibnamefont {Vignolle}}, \bibinfo
  {author} {\bibfnamefont {O.~J.}\ \bibnamefont {Lipscombe}}, \bibinfo {author}
  {\bibfnamefont {S.~M.}\ \bibnamefont {Hayden}}, \bibinfo {author}
  {\bibfnamefont {Y.}~\bibnamefont {Tanabe}}, \bibinfo {author} {\bibfnamefont
  {T.}~\bibnamefont {Adachi}}, \bibinfo {author} {\bibfnamefont
  {Y.}~\bibnamefont {Koike}}, \bibinfo {author} {\bibfnamefont
  {M.}~\bibnamefont {Nohara}}, \bibinfo {author} {\bibfnamefont
  {H.}~\bibnamefont {Takagi}}, \bibinfo {author} {\bibfnamefont
  {C.}~\bibnamefont {Proust}}, \ and\ \bibinfo {author} {\bibfnamefont {N.~E.}\
  \bibnamefont {Hussey}},\ }\href {\doibase 10.1126/science.1165015} {\bibfield
   {journal} {\bibinfo  {journal} {Science}\ }\textbf {\bibinfo {volume}
  {323}},\ \bibinfo {pages} {603} (\bibinfo {year} {2009})}\BibitemShut
  {NoStop}%
\bibitem [{\citenamefont {Aoki}\ \emph {et~al.}(2013)\citenamefont {Aoki},
  \citenamefont {Knafo},\ and\ \citenamefont {Sheikin}}]{aoki13}%
  \BibitemOpen
  \bibfield  {author} {\bibinfo {author} {\bibfnamefont {D.}~\bibnamefont
  {Aoki}}, \bibinfo {author} {\bibfnamefont {W.}~\bibnamefont {Knafo}}, \ and\
  \bibinfo {author} {\bibfnamefont {I.}~\bibnamefont {Sheikin}},\ }\href@noop
  {} {\bibfield  {journal} {\bibinfo  {journal} {C. R. Physique}\ }\textbf
  {\bibinfo {volume} {14}},\ \bibinfo {pages} {53} (\bibinfo {year}
  {2013})}\BibitemShut {NoStop}%
\bibitem [{\citenamefont {Settai}\ \emph {et~al.}(2007)\citenamefont {Settai},
  \citenamefont {Takeuchi},\ and\ \citenamefont {\={O}nuki}}]{settai07}%
  \BibitemOpen
  \bibfield  {author} {\bibinfo {author} {\bibfnamefont {R.}~\bibnamefont
  {Settai}}, \bibinfo {author} {\bibfnamefont {T.}~\bibnamefont {Takeuchi}}, \
  and\ \bibinfo {author} {\bibfnamefont {Y.}~\bibnamefont {\={O}nuki}},\ }\href
  {\doibase 10.1143/JPSJ.76.051003} {\bibfield  {journal} {\bibinfo  {journal}
  {Journal of the Physical Society of Japan}\ }\textbf {\bibinfo {volume}
  {76}},\ \bibinfo {pages} {051003} (\bibinfo {year} {2007})}\BibitemShut
  {NoStop}%
\bibitem [{\citenamefont {Lee}\ \emph {et~al.}(1986)\citenamefont {Lee},
  \citenamefont {Rice}, \citenamefont {Serene}, \citenamefont {Sham},\ and\
  \citenamefont {Wilkins}}]{lee86}%
  \BibitemOpen
  \bibfield  {author} {\bibinfo {author} {\bibfnamefont {P.~A.}\ \bibnamefont
  {Lee}}, \bibinfo {author} {\bibfnamefont {T.~M.}\ \bibnamefont {Rice}},
  \bibinfo {author} {\bibfnamefont {J.~W.}\ \bibnamefont {Serene}}, \bibinfo
  {author} {\bibfnamefont {L.~J.}\ \bibnamefont {Sham}}, \ and\ \bibinfo
  {author} {\bibfnamefont {J.~W.}\ \bibnamefont {Wilkins}},\ }\href@noop {}
  {\bibfield  {journal} {\bibinfo  {journal} {Comment. Cond. Mat. Phys.}\
  }\textbf {\bibinfo {volume} {12}},\ \bibinfo {pages} {99} (\bibinfo {year}
  {1986})}\BibitemShut {NoStop}%
\bibitem [{\citenamefont {Ohashi}\ \emph {et~al.}(2003)\citenamefont {Ohashi},
  \citenamefont {Oomi}, \citenamefont {Koiwai}, \citenamefont {Hedo},\ and\
  \citenamefont {Uwatoko}}]{ohashi03}%
  \BibitemOpen
  \bibfield  {author} {\bibinfo {author} {\bibfnamefont {M.}~\bibnamefont
  {Ohashi}}, \bibinfo {author} {\bibfnamefont {G.}~\bibnamefont {Oomi}},
  \bibinfo {author} {\bibfnamefont {S.}~\bibnamefont {Koiwai}}, \bibinfo
  {author} {\bibfnamefont {M.}~\bibnamefont {Hedo}}, \ and\ \bibinfo {author}
  {\bibfnamefont {Y.}~\bibnamefont {Uwatoko}},\ }\href@noop {} {\bibfield
  {journal} {\bibinfo  {journal} {Phys. Rev. B}\ }\textbf {\bibinfo {volume}
  {68}},\ \bibinfo {pages} {144428} (\bibinfo {year} {2003})}\BibitemShut
  {NoStop}%
\bibitem [{\citenamefont {Scheerer}\ \emph {et~al.}(2012)\citenamefont
  {Scheerer}, \citenamefont {Knafo}, \citenamefont {Aoki}, \citenamefont
  {Ballon}, \citenamefont {Mari}, \citenamefont {Vignolles},\ and\
  \citenamefont {Flouquet}}]{scheerer12}%
  \BibitemOpen
  \bibfield  {author} {\bibinfo {author} {\bibfnamefont {G.~W.}\ \bibnamefont
  {Scheerer}}, \bibinfo {author} {\bibfnamefont {W.}~\bibnamefont {Knafo}},
  \bibinfo {author} {\bibfnamefont {D.}~\bibnamefont {Aoki}}, \bibinfo {author}
  {\bibfnamefont {G.}~\bibnamefont {Ballon}}, \bibinfo {author} {\bibfnamefont
  {A.}~\bibnamefont {Mari}}, \bibinfo {author} {\bibfnamefont {D.}~\bibnamefont
  {Vignolles}}, \ and\ \bibinfo {author} {\bibfnamefont {J.}~\bibnamefont
  {Flouquet}},\ }\href@noop {} {\bibfield  {journal} {\bibinfo  {journal}
  {Phys. Rev. B}\ }\textbf {\bibinfo {volume} {85}},\ \bibinfo {pages} {094402}
  (\bibinfo {year} {2012})}\BibitemShut {NoStop}%
\bibitem [{\citenamefont {Tsujii}\ \emph {et~al.}(2005)\citenamefont {Tsujii},
  \citenamefont {Kontani},\ and\ \citenamefont {Yoshimura}}]{tsujii2005}%
  \BibitemOpen
  \bibfield  {author} {\bibinfo {author} {\bibfnamefont {N.}~\bibnamefont
  {Tsujii}}, \bibinfo {author} {\bibfnamefont {H.}~\bibnamefont {Kontani}}, \
  and\ \bibinfo {author} {\bibfnamefont {K.}~\bibnamefont {Yoshimura}},\
  }\href@noop {} {\bibfield  {journal} {\bibinfo  {journal} {Phys. Rev. Lett.}\
  }\textbf {\bibinfo {volume} {94}},\ \bibinfo {pages} {057201} (\bibinfo
  {year} {2005})}\BibitemShut {NoStop}%
\bibitem [{\citenamefont {Daou}\ \emph {et~al.}(2006)\citenamefont {Daou},
  \citenamefont {Bergemann},\ and\ \citenamefont {Julian}}]{daou06}%
  \BibitemOpen
  \bibfield  {author} {\bibinfo {author} {\bibfnamefont {R.}~\bibnamefont
  {Daou}}, \bibinfo {author} {\bibfnamefont {C.}~\bibnamefont {Bergemann}}, \
  and\ \bibinfo {author} {\bibfnamefont {S.~R.}\ \bibnamefont {Julian}},\
  }\href@noop {} {\bibfield  {journal} {\bibinfo  {journal} {Phys. Rev. Lett.}\
  }\textbf {\bibinfo {volume} {96}},\ \bibinfo {pages} {026401} (\bibinfo
  {year} {2006})}\BibitemShut {NoStop}%
\bibitem [{\citenamefont {Fisher}\ \emph {et~al.}(1991)\citenamefont {Fisher},
  \citenamefont {Marcenat}, \citenamefont {Phillips}, \citenamefont {Haen},
  \citenamefont {Lapierre}, \citenamefont {Lejay}, \citenamefont {Flouquet},\
  and\ \citenamefont {Voiron}}]{fisher91}%
  \BibitemOpen
  \bibfield  {author} {\bibinfo {author} {\bibfnamefont {R.~A.}\ \bibnamefont
  {Fisher}}, \bibinfo {author} {\bibfnamefont {C.}~\bibnamefont {Marcenat}},
  \bibinfo {author} {\bibfnamefont {N.~E.}\ \bibnamefont {Phillips}}, \bibinfo
  {author} {\bibfnamefont {P.}~\bibnamefont {Haen}}, \bibinfo {author}
  {\bibfnamefont {F.}~\bibnamefont {Lapierre}}, \bibinfo {author}
  {\bibfnamefont {P.}~\bibnamefont {Lejay}}, \bibinfo {author} {\bibfnamefont
  {J.}~\bibnamefont {Flouquet}}, \ and\ \bibinfo {author} {\bibfnamefont
  {J.}~\bibnamefont {Voiron}},\ }\href {\doibase 10.1007/BF00681617} {\bibfield
   {journal} {\bibinfo  {journal} {Journal of Low Temperature Physics}\
  }\textbf {\bibinfo {volume} {84}},\ \bibinfo {pages} {49} (\bibinfo {year}
  {1991})}\BibitemShut {NoStop}%
\bibitem [{\citenamefont {Gegenwart}\ \emph {et~al.}(2008)\citenamefont
  {Gegenwart}, \citenamefont {Si},\ and\ \citenamefont
  {Steglich}}]{gegenwart08}%
  \BibitemOpen
  \bibfield  {author} {\bibinfo {author} {\bibfnamefont {P.}~\bibnamefont
  {Gegenwart}}, \bibinfo {author} {\bibfnamefont {Q.}~\bibnamefont {Si}}, \
  and\ \bibinfo {author} {\bibfnamefont {F.}~\bibnamefont {Steglich}},\ }\href
  {\doibase 10.1038/nphys892} {\bibfield  {journal} {\bibinfo  {journal} {Nat.
  Phys.}\ }\textbf {\bibinfo {volume} {4}},\ \bibinfo {pages} {186} (\bibinfo
  {year} {2008})}\BibitemShut {NoStop}%
\bibitem [{\citenamefont {Flouquet}(2005)}]{flouquet05}%
  \BibitemOpen
  \bibfield  {author} {\bibinfo {author} {\bibfnamefont {J.}~\bibnamefont
  {Flouquet}},\ }\href@noop {} {\bibfield  {journal} {\bibinfo  {journal}
  {Prog. Low Temp. Phys.}\ }\textbf {\bibinfo {volume} {15}},\ \bibinfo {pages}
  {139} (\bibinfo {year} {2005})}\BibitemShut {NoStop}%
\bibitem [{\citenamefont {L\"ohneysen}\ \emph {et~al.}(2007)\citenamefont
  {L\"ohneysen}, \citenamefont {Rosch}, \citenamefont {Vojta},\ and\
  \citenamefont {W\"olfle}}]{lohneysen07}%
  \BibitemOpen
  \bibfield  {author} {\bibinfo {author} {\bibfnamefont {H.~v.}\ \bibnamefont
  {L\"ohneysen}}, \bibinfo {author} {\bibfnamefont {A.}~\bibnamefont {Rosch}},
  \bibinfo {author} {\bibfnamefont {M.}~\bibnamefont {Vojta}}, \ and\ \bibinfo
  {author} {\bibfnamefont {P.}~\bibnamefont {W\"olfle}},\ }\href {\doibase
  10.1103/RevModPhys.79.1015} {\bibfield  {journal} {\bibinfo  {journal} {Rev.
  Mod. Phys.}\ }\textbf {\bibinfo {volume} {79}},\ \bibinfo {pages} {1015}
  (\bibinfo {year} {2007})}\BibitemShut {NoStop}%
\bibitem [{\citenamefont {Pfleiderer}(2009)}]{pfleiderer09}%
  \BibitemOpen
  \bibfield  {author} {\bibinfo {author} {\bibfnamefont {C.}~\bibnamefont
  {Pfleiderer}},\ }\href {\doibase 10.1103/RevModPhys.81.1551} {\bibfield
  {journal} {\bibinfo  {journal} {Rev. Mod. Phys.}\ }\textbf {\bibinfo {volume}
  {81}},\ \bibinfo {pages} {1551} (\bibinfo {year} {2009})}\BibitemShut
  {NoStop}%
\bibitem [{\citenamefont {Friedemann}\ \emph {et~al.}(2009)\citenamefont
  {Friedemann}, \citenamefont {Westerkamp}, \citenamefont {Brando},
  \citenamefont {Oeschler}, \citenamefont {Wirth}, \citenamefont {Gegenwart},
  \citenamefont {Krellner}, \citenamefont {Geibel},\ and\ \citenamefont
  {Steglich}}]{friedemann09}%
  \BibitemOpen
  \bibfield  {author} {\bibinfo {author} {\bibfnamefont {S.}~\bibnamefont
  {Friedemann}}, \bibinfo {author} {\bibfnamefont {T.}~\bibnamefont
  {Westerkamp}}, \bibinfo {author} {\bibfnamefont {M.}~\bibnamefont {Brando}},
  \bibinfo {author} {\bibfnamefont {N.}~\bibnamefont {Oeschler}}, \bibinfo
  {author} {\bibfnamefont {S.}~\bibnamefont {Wirth}}, \bibinfo {author}
  {\bibfnamefont {P.}~\bibnamefont {Gegenwart}}, \bibinfo {author}
  {\bibfnamefont {C.}~\bibnamefont {Krellner}}, \bibinfo {author}
  {\bibfnamefont {C.}~\bibnamefont {Geibel}}, \ and\ \bibinfo {author}
  {\bibfnamefont {F.}~\bibnamefont {Steglich}},\ }\href@noop {} {\bibfield
  {journal} {\bibinfo  {journal} {Nat. Phys.}\ }\textbf {\bibinfo {volume}
  {5}},\ \bibinfo {pages} {465 } (\bibinfo {year} {2009})}\BibitemShut
  {NoStop}%
\bibitem [{not({\natexlab{a}})}]{note_SF}%
  \BibitemOpen
  \bibinfo {note} {A spin-flop transition corresponds to the sudden alignment
  of the AF moments perpendicular to the magnetic field and is accompanied by a
  step-like variation of the magnetization.}\BibitemShut {Stop}%
\bibitem [{\citenamefont {Grube}\ \emph {et~al.}(2007)\citenamefont {Grube},
  \citenamefont {Knafo}, \citenamefont {Drobnik}, \citenamefont {Adelmann},
  \citenamefont {Wolf}, \citenamefont {Meingast},\ and\ \citenamefont
  {Löhneysen}}]{grube07}%
  \BibitemOpen
  \bibfield  {author} {\bibinfo {author} {\bibfnamefont {K.}~\bibnamefont
  {Grube}}, \bibinfo {author} {\bibfnamefont {W.}~\bibnamefont {Knafo}},
  \bibinfo {author} {\bibfnamefont {S.}~\bibnamefont {Drobnik}}, \bibinfo
  {author} {\bibfnamefont {P.}~\bibnamefont {Adelmann}}, \bibinfo {author}
  {\bibfnamefont {T.}~\bibnamefont {Wolf}}, \bibinfo {author} {\bibfnamefont
  {C.}~\bibnamefont {Meingast}}, \ and\ \bibinfo {author} {\bibfnamefont
  {H.}~\bibnamefont {Löhneysen}},\ }\href@noop {} {\bibfield  {journal}
  {\bibinfo  {journal} {Journal of Magnetism and Magnetic Materials}\ }\textbf
  {\bibinfo {volume} {310}},\ \bibinfo {pages} {354 } (\bibinfo {year}
  {2007})}\BibitemShut {NoStop}%
\bibitem [{\citenamefont {Knafo}\ \emph {et~al.}(2007)\citenamefont {Knafo},
  \citenamefont {Meingast}, \citenamefont {Grube}, \citenamefont {Drobnik},
  \citenamefont {Popovich}, \citenamefont {Schweiss}, \citenamefont {Adelmann},
  \citenamefont {Wolf},\ and\ \citenamefont {v.~L\"ohneysen}}]{knafo07}%
  \BibitemOpen
  \bibfield  {author} {\bibinfo {author} {\bibfnamefont {W.}~\bibnamefont
  {Knafo}}, \bibinfo {author} {\bibfnamefont {C.}~\bibnamefont {Meingast}},
  \bibinfo {author} {\bibfnamefont {K.}~\bibnamefont {Grube}}, \bibinfo
  {author} {\bibfnamefont {S.}~\bibnamefont {Drobnik}}, \bibinfo {author}
  {\bibfnamefont {P.}~\bibnamefont {Popovich}}, \bibinfo {author}
  {\bibfnamefont {P.}~\bibnamefont {Schweiss}}, \bibinfo {author}
  {\bibfnamefont {P.}~\bibnamefont {Adelmann}}, \bibinfo {author}
  {\bibfnamefont {T.}~\bibnamefont {Wolf}}, \ and\ \bibinfo {author}
  {\bibfnamefont {H.}~\bibnamefont {v.~L\"ohneysen}},\ }\href@noop {}
  {\bibfield  {journal} {\bibinfo  {journal} {Phys. Rev. Lett.}\ }\textbf
  {\bibinfo {volume} {99}},\ \bibinfo {pages} {137206} (\bibinfo {year}
  {2007})}\BibitemShut {NoStop}%
\bibitem [{\citenamefont {Stryjewski}\ and\ \citenamefont
  {Giordano}(1977)}]{stryjewski05}%
  \BibitemOpen
  \bibfield  {author} {\bibinfo {author} {\bibfnamefont {E.}~\bibnamefont
  {Stryjewski}}\ and\ \bibinfo {author} {\bibfnamefont {N.}~\bibnamefont
  {Giordano}},\ }\href@noop {} {\bibfield  {journal} {\bibinfo  {journal} {Adv.
  Phys.}\ }\textbf {\bibinfo {volume} {26}},\ \bibinfo {pages} {487–650}
  (\bibinfo {year} {1977})}\BibitemShut {NoStop}%
\bibitem [{\citenamefont {Settai}\ \emph {et~al.}(1997)\citenamefont {Settai},
  \citenamefont {Misawa}, \citenamefont {Araki}, \citenamefont {Kosaki},
  \citenamefont {Sugiyama}, \citenamefont {Takeuchi}, \citenamefont {Kindo},
  \citenamefont {Haga}, \citenamefont {Yamamoto},\ and\ \citenamefont
  {\={O}nuki}}]{settai97}%
  \BibitemOpen
  \bibfield  {author} {\bibinfo {author} {\bibfnamefont {R.}~\bibnamefont
  {Settai}}, \bibinfo {author} {\bibfnamefont {A.}~\bibnamefont {Misawa}},
  \bibinfo {author} {\bibfnamefont {S.}~\bibnamefont {Araki}}, \bibinfo
  {author} {\bibfnamefont {M.}~\bibnamefont {Kosaki}}, \bibinfo {author}
  {\bibfnamefont {K.}~\bibnamefont {Sugiyama}}, \bibinfo {author}
  {\bibfnamefont {T.}~\bibnamefont {Takeuchi}}, \bibinfo {author}
  {\bibfnamefont {K.}~\bibnamefont {Kindo}}, \bibinfo {author} {\bibfnamefont
  {Y.}~\bibnamefont {Haga}}, \bibinfo {author} {\bibfnamefont {E.}~\bibnamefont
  {Yamamoto}}, \ and\ \bibinfo {author} {\bibfnamefont {Y.}~\bibnamefont
  {\={O}nuki}},\ }\href@noop {} {\bibfield  {journal} {\bibinfo  {journal}
  {Prog. Low Temp. Phys.}\ }\textbf {\bibinfo {volume} {66}},\ \bibinfo {pages}
  {2260} (\bibinfo {year} {1997})}\BibitemShut {NoStop}%
\bibitem [{\citenamefont {Knafo}\ \emph {et~al.}(2010)\citenamefont {Knafo},
  \citenamefont {Aoki}, \citenamefont {Vignolles}, \citenamefont {Vignolle},
  \citenamefont {Klein}, \citenamefont {Jaudet}, \citenamefont {Villaume},
  \citenamefont {Proust},\ and\ \citenamefont {Flouquet}}]{knafo10}%
  \BibitemOpen
  \bibfield  {author} {\bibinfo {author} {\bibfnamefont {W.}~\bibnamefont
  {Knafo}}, \bibinfo {author} {\bibfnamefont {D.}~\bibnamefont {Aoki}},
  \bibinfo {author} {\bibfnamefont {D.}~\bibnamefont {Vignolles}}, \bibinfo
  {author} {\bibfnamefont {B.}~\bibnamefont {Vignolle}}, \bibinfo {author}
  {\bibfnamefont {Y.}~\bibnamefont {Klein}}, \bibinfo {author} {\bibfnamefont
  {C.}~\bibnamefont {Jaudet}}, \bibinfo {author} {\bibfnamefont
  {A.}~\bibnamefont {Villaume}}, \bibinfo {author} {\bibfnamefont
  {C.}~\bibnamefont {Proust}}, \ and\ \bibinfo {author} {\bibfnamefont
  {J.}~\bibnamefont {Flouquet}},\ }\href {\doibase 10.1103/PhysRevB.81.094403}
  {\bibfield  {journal} {\bibinfo  {journal} {Phys. Rev. B}\ }\textbf {\bibinfo
  {volume} {81}},\ \bibinfo {pages} {094403} (\bibinfo {year}
  {2010})}\BibitemShut {NoStop}%
\bibitem [{not({\natexlab{b}})}]{note_Hc_Hm}%
  \BibitemOpen
  \bibinfo {note} {By tradition in the heavy-fermion community, the
  metamagnetic transition at the border of antiferromagnetism is usually noted
  $H_c$, while the pseudo-metamagnetic crossover at the border of correlated
  paramagnetism is usually noted $H_m$.}\BibitemShut {Stop}%
\bibitem [{\citenamefont {Rossat-Mignod}\ \emph {et~al.}(1988)\citenamefont
  {Rossat-Mignod}, \citenamefont {Regnault}, \citenamefont {Jacoud},
  \citenamefont {Vettier}, \citenamefont {Lejay}, \citenamefont {Flouquet},
  \citenamefont {Walker}, \citenamefont {Jaccard},\ and\ \citenamefont
  {Amato}}]{rossat88}%
  \BibitemOpen
  \bibfield  {author} {\bibinfo {author} {\bibfnamefont {J.}~\bibnamefont
  {Rossat-Mignod}}, \bibinfo {author} {\bibfnamefont {L.}~\bibnamefont
  {Regnault}}, \bibinfo {author} {\bibfnamefont {J.}~\bibnamefont {Jacoud}},
  \bibinfo {author} {\bibfnamefont {C.}~\bibnamefont {Vettier}}, \bibinfo
  {author} {\bibfnamefont {P.}~\bibnamefont {Lejay}}, \bibinfo {author}
  {\bibfnamefont {J.}~\bibnamefont {Flouquet}}, \bibinfo {author}
  {\bibfnamefont {E.}~\bibnamefont {Walker}}, \bibinfo {author} {\bibfnamefont
  {D.}~\bibnamefont {Jaccard}}, \ and\ \bibinfo {author} {\bibfnamefont
  {A.}~\bibnamefont {Amato}},\ }\href@noop {} {\bibfield  {journal} {\bibinfo
  {journal} {J. Magn. Magn. Mater.}\ }\textbf {\bibinfo {volume} {76-77}},\
  \bibinfo {pages} {376} (\bibinfo {year} {1988})}\BibitemShut {NoStop}%
\bibitem [{\citenamefont {Flouquet}\ \emph {et~al.}(2004)\citenamefont
  {Flouquet}, \citenamefont {Haga}, \citenamefont {Haen}, \citenamefont
  {Braithwaite}, \citenamefont {Knebel}, \citenamefont {Raymond},\ and\
  \citenamefont {Kambe}}]{flouquet04}%
  \BibitemOpen
  \bibfield  {author} {\bibinfo {author} {\bibfnamefont {J.}~\bibnamefont
  {Flouquet}}, \bibinfo {author} {\bibfnamefont {Y.}~\bibnamefont {Haga}},
  \bibinfo {author} {\bibfnamefont {P.}~\bibnamefont {Haen}}, \bibinfo {author}
  {\bibfnamefont {D.}~\bibnamefont {Braithwaite}}, \bibinfo {author}
  {\bibfnamefont {G.}~\bibnamefont {Knebel}}, \bibinfo {author} {\bibfnamefont
  {S.}~\bibnamefont {Raymond}}, \ and\ \bibinfo {author} {\bibfnamefont
  {S.}~\bibnamefont {Kambe}},\ }\href@noop {} {\bibfield  {journal} {\bibinfo
  {journal} {J. Magn. Magn. Mater.}\ }\textbf {\bibinfo {volume} {272-276}},\
  \bibinfo {pages} {27} (\bibinfo {year} {2004})}\BibitemShut {NoStop}%
\bibitem [{\citenamefont {Grier}\ \emph {et~al.}(1984)\citenamefont {Grier},
  \citenamefont {Lawrence}, \citenamefont {Murgai},\ and\ \citenamefont
  {Parks}}]{grier84}%
  \BibitemOpen
  \bibfield  {author} {\bibinfo {author} {\bibfnamefont {B.}~\bibnamefont
  {Grier}}, \bibinfo {author} {\bibfnamefont {J.}~\bibnamefont {Lawrence}},
  \bibinfo {author} {\bibfnamefont {V.}~\bibnamefont {Murgai}}, \ and\ \bibinfo
  {author} {\bibfnamefont {R.}~\bibnamefont {Parks}},\ }\href@noop {}
  {\bibfield  {journal} {\bibinfo  {journal} {Phys. Rev. B}\ }\textbf {\bibinfo
  {volume} {29}},\ \bibinfo {pages} {2664} (\bibinfo {year}
  {1984})}\BibitemShut {NoStop}%
\bibitem [{\citenamefont {Kawarazaki}\ \emph {et~al.}(2000)\citenamefont
  {Kawarazaki}, \citenamefont {Sato}, \citenamefont {Miyako}, \citenamefont
  {Chigusa}, \citenamefont {Watanabe}, \citenamefont {Metoki}, \citenamefont
  {Koike},\ and\ \citenamefont {Nishi}}]{kawarasaki00}%
  \BibitemOpen
  \bibfield  {author} {\bibinfo {author} {\bibfnamefont {S.}~\bibnamefont
  {Kawarazaki}}, \bibinfo {author} {\bibfnamefont {M.}~\bibnamefont {Sato}},
  \bibinfo {author} {\bibfnamefont {Y.}~\bibnamefont {Miyako}}, \bibinfo
  {author} {\bibfnamefont {N.}~\bibnamefont {Chigusa}}, \bibinfo {author}
  {\bibfnamefont {K.}~\bibnamefont {Watanabe}}, \bibinfo {author}
  {\bibfnamefont {N.}~\bibnamefont {Metoki}}, \bibinfo {author} {\bibfnamefont
  {Y.}~\bibnamefont {Koike}}, \ and\ \bibinfo {author} {\bibfnamefont
  {M.}~\bibnamefont {Nishi}},\ }\href@noop {} {\bibfield  {journal} {\bibinfo
  {journal} {Phys. Rev. B}\ }\textbf {\bibinfo {volume} {61}},\ \bibinfo
  {pages} {4167} (\bibinfo {year} {2000})}\BibitemShut {NoStop}%
\bibitem [{\citenamefont {Muramatsu}\ \emph {et~al.}(1999)\citenamefont
  {Muramatsu}, \citenamefont {Eda}, \citenamefont {Kobayashi}, \citenamefont
  {Eremets}, \citenamefont {Amaya}, \citenamefont {Araki}, \citenamefont
  {Settai},\ and\ \citenamefont {\={O}nuki}}]{muramatsu99}%
  \BibitemOpen
  \bibfield  {author} {\bibinfo {author} {\bibfnamefont {T.}~\bibnamefont
  {Muramatsu}}, \bibinfo {author} {\bibfnamefont {S.}~\bibnamefont {Eda}},
  \bibinfo {author} {\bibfnamefont {T.}~\bibnamefont {Kobayashi}}, \bibinfo
  {author} {\bibfnamefont {M.}~\bibnamefont {Eremets}}, \bibinfo {author}
  {\bibfnamefont {K.}~\bibnamefont {Amaya}}, \bibinfo {author} {\bibfnamefont
  {S.}~\bibnamefont {Araki}}, \bibinfo {author} {\bibfnamefont
  {R.}~\bibnamefont {Settai}}, \ and\ \bibinfo {author} {\bibfnamefont
  {Y.}~\bibnamefont {\={O}nuki}},\ }\href@noop {} {\bibfield  {journal}
  {\bibinfo  {journal} {Physica B}\ }\textbf {\bibinfo {volume} {259--261}},\
  \bibinfo {pages} {61} (\bibinfo {year} {1999})}\BibitemShut {NoStop}%
\bibitem [{\citenamefont {Movshovich}\ \emph {et~al.}(1996)\citenamefont
  {Movshovich}, \citenamefont {Graf}, \citenamefont {Mandrus}, \citenamefont
  {Thompson}, \citenamefont {Smith},\ and\ \citenamefont
  {Fisk}}]{movshovich96}%
  \BibitemOpen
  \bibfield  {author} {\bibinfo {author} {\bibfnamefont {R.}~\bibnamefont
  {Movshovich}}, \bibinfo {author} {\bibfnamefont {T.}~\bibnamefont {Graf}},
  \bibinfo {author} {\bibfnamefont {D.}~\bibnamefont {Mandrus}}, \bibinfo
  {author} {\bibfnamefont {J.~D.}\ \bibnamefont {Thompson}}, \bibinfo {author}
  {\bibfnamefont {J.~L.}\ \bibnamefont {Smith}}, \ and\ \bibinfo {author}
  {\bibfnamefont {Z.}~\bibnamefont {Fisk}},\ }\href@noop {} {\bibfield
  {journal} {\bibinfo  {journal} {Phys. Rev. B}\ }\textbf {\bibinfo {volume}
  {53}},\ \bibinfo {pages} {8241} (\bibinfo {year} {1996})}\BibitemShut
  {NoStop}%
\bibitem [{\citenamefont {Araki}\ \emph {et~al.}(2002)\citenamefont {Araki},
  \citenamefont {Nakashima}, \citenamefont {Settai}, \citenamefont
  {Kobayashi},\ and\ \citenamefont {\={O}nuki}}]{araki02}%
  \BibitemOpen
  \bibfield  {author} {\bibinfo {author} {\bibfnamefont {S.}~\bibnamefont
  {Araki}}, \bibinfo {author} {\bibfnamefont {M.}~\bibnamefont {Nakashima}},
  \bibinfo {author} {\bibfnamefont {R.}~\bibnamefont {Settai}}, \bibinfo
  {author} {\bibfnamefont {T.~C.}\ \bibnamefont {Kobayashi}}, \ and\ \bibinfo
  {author} {\bibfnamefont {Y.}~\bibnamefont {\={O}nuki}},\ }\href
  {http://stacks.iop.org/0953-8984/14/i=21/a=102} {\bibfield  {journal}
  {\bibinfo  {journal} {Journal of Physics: Condensed Matter}\ }\textbf
  {\bibinfo {volume} {14}},\ \bibinfo {pages} {L377} (\bibinfo {year}
  {2002})}\BibitemShut {NoStop}%
\bibitem [{\citenamefont {Abe}\ \emph {et~al.}(1997)\citenamefont {Abe},
  \citenamefont {Suzuki}, \citenamefont {Kitazawa}, \citenamefont {Matsumo},\
  and\ \citenamefont {Kido}}]{abe97}%
  \BibitemOpen
  \bibfield  {author} {\bibinfo {author} {\bibfnamefont {H.}~\bibnamefont
  {Abe}}, \bibinfo {author} {\bibfnamefont {H.}~\bibnamefont {Suzuki}},
  \bibinfo {author} {\bibfnamefont {H.}~\bibnamefont {Kitazawa}}, \bibinfo
  {author} {\bibfnamefont {T.}~\bibnamefont {Matsumo}}, \ and\ \bibinfo
  {author} {\bibfnamefont {G.}~\bibnamefont {Kido}},\ }\href {\doibase
  10.1143/JPSJ.66.2525} {\bibfield  {journal} {\bibinfo  {journal} {Journal of
  the Physical Society of Japan}\ }\textbf {\bibinfo {volume} {66}},\ \bibinfo
  {pages} {2525} (\bibinfo {year} {1997})}\BibitemShut {NoStop}%
\bibitem [{\citenamefont {Hamamoto}\ \emph {et~al.}(2000)\citenamefont
  {Hamamoto}, \citenamefont {Kindo}, \citenamefont {Kobayashi}, \citenamefont
  {Uwatoko}, \citenamefont {Araki}, \citenamefont {Settai},\ and\ \citenamefont
  {\={O}nuki}}]{hamamoto00}%
  \BibitemOpen
  \bibfield  {author} {\bibinfo {author} {\bibfnamefont {T.}~\bibnamefont
  {Hamamoto}}, \bibinfo {author} {\bibfnamefont {K.}~\bibnamefont {Kindo}},
  \bibinfo {author} {\bibfnamefont {T.}~\bibnamefont {Kobayashi}}, \bibinfo
  {author} {\bibfnamefont {Y.}~\bibnamefont {Uwatoko}}, \bibinfo {author}
  {\bibfnamefont {S.}~\bibnamefont {Araki}}, \bibinfo {author} {\bibfnamefont
  {R.}~\bibnamefont {Settai}}, \ and\ \bibinfo {author} {\bibfnamefont
  {Y.}~\bibnamefont {\={O}nuki}},\ }\href {\doibase
  http://dx.doi.org/10.1016/S0921-4526(99)00973-4} {\bibfield  {journal}
  {\bibinfo  {journal} {Physica B: Condensed Matter}\ }\textbf {\bibinfo
  {volume} {281–282}},\ \bibinfo {pages} {64 } (\bibinfo {year}
  {2000})}\BibitemShut {NoStop}%
\bibitem [{\citenamefont {Knebel}\ \emph {et~al.}(2011)\citenamefont {Knebel},
  \citenamefont {Buhot}, \citenamefont {Aoki}, \citenamefont {Lapertot},
  \citenamefont {Raymond}, \citenamefont {Ressouche},\ and\ \citenamefont
  {Flouquet}}]{knebel11}%
  \BibitemOpen
  \bibfield  {author} {\bibinfo {author} {\bibfnamefont {G.}~\bibnamefont
  {Knebel}}, \bibinfo {author} {\bibfnamefont {J.}~\bibnamefont {Buhot}},
  \bibinfo {author} {\bibfnamefont {D.}~\bibnamefont {Aoki}}, \bibinfo {author}
  {\bibfnamefont {G.}~\bibnamefont {Lapertot}}, \bibinfo {author}
  {\bibfnamefont {S.}~\bibnamefont {Raymond}}, \bibinfo {author} {\bibfnamefont
  {E.}~\bibnamefont {Ressouche}}, \ and\ \bibinfo {author} {\bibfnamefont
  {J.}~\bibnamefont {Flouquet}},\ }\href@noop {} {\bibfield  {journal}
  {\bibinfo  {journal} {J. Phys. Soc. Jpn.}\ }\textbf {\bibinfo {volume}
  {80}},\ \bibinfo {pages} {SA001} (\bibinfo {year} {2011})}\BibitemShut
  {NoStop}%
\bibitem [{\citenamefont {Zaum}\ \emph {et~al.}(2011)\citenamefont {Zaum},
  \citenamefont {Grube}, \citenamefont {Schäfer}, \citenamefont {Bauer},
  \citenamefont {Thompson},\ and\ \citenamefont {v.~Löhneysen}}]{zaum05}%
  \BibitemOpen
  \bibfield  {author} {\bibinfo {author} {\bibfnamefont {S.}~\bibnamefont
  {Zaum}}, \bibinfo {author} {\bibfnamefont {K.}~\bibnamefont {Grube}},
  \bibinfo {author} {\bibfnamefont {R.}~\bibnamefont {Schäfer}}, \bibinfo
  {author} {\bibfnamefont {E.}~\bibnamefont {Bauer}}, \bibinfo {author}
  {\bibfnamefont {J.}~\bibnamefont {Thompson}}, \ and\ \bibinfo {author}
  {\bibfnamefont {H.}~\bibnamefont {v.~Löhneysen}},\ }\href@noop {} {\bibfield
   {journal} {\bibinfo  {journal} {Phys. Rev. Lett.}\ }\textbf {\bibinfo
  {volume} {106}},\ \bibinfo {pages} {087003} (\bibinfo {year}
  {2011})}\BibitemShut {NoStop}%
\bibitem [{\citenamefont {Matsubayashi}\ \emph {et~al.}(2015)\citenamefont
  {Matsubayashi}, \citenamefont {Hirayama}, \citenamefont {Yamashita},
  \citenamefont {Ohara}, \citenamefont {Kawamura}, \citenamefont {Mizumaki},
  \citenamefont {Ishimatsu}, \citenamefont {Watanabe}, \citenamefont
  {Kitagawa},\ and\ \citenamefont {Uwatoko}}]{matsubayashi15}%
  \BibitemOpen
  \bibfield  {author} {\bibinfo {author} {\bibfnamefont {K.}~\bibnamefont
  {Matsubayashi}}, \bibinfo {author} {\bibfnamefont {T.}~\bibnamefont
  {Hirayama}}, \bibinfo {author} {\bibfnamefont {T.}~\bibnamefont {Yamashita}},
  \bibinfo {author} {\bibfnamefont {S.}~\bibnamefont {Ohara}}, \bibinfo
  {author} {\bibfnamefont {N.}~\bibnamefont {Kawamura}}, \bibinfo {author}
  {\bibfnamefont {M.}~\bibnamefont {Mizumaki}}, \bibinfo {author}
  {\bibfnamefont {N.}~\bibnamefont {Ishimatsu}}, \bibinfo {author}
  {\bibfnamefont {S.}~\bibnamefont {Watanabe}}, \bibinfo {author}
  {\bibfnamefont {K.}~\bibnamefont {Kitagawa}}, \ and\ \bibinfo {author}
  {\bibfnamefont {Y.}~\bibnamefont {Uwatoko}},\ }\href@noop {} {\bibfield
  {journal} {\bibinfo  {journal} {Phys. Rev. Lett.}\ }\textbf {\bibinfo
  {volume} {114}},\ \bibinfo {pages} {086401} (\bibinfo {year}
  {2015})}\BibitemShut {NoStop}%
\bibitem [{\citenamefont {Ohashi}\ \emph {et~al.}(2002)\citenamefont {Ohashi},
  \citenamefont {Honda}, \citenamefont {Eto}, \citenamefont {Kaji},
  \citenamefont {Minamitake}, \citenamefont {Oomi}, \citenamefont {Koiwai},\
  and\ \citenamefont {Uwatoko}}]{ohashi02}%
  \BibitemOpen
  \bibfield  {author} {\bibinfo {author} {\bibfnamefont {M.}~\bibnamefont
  {Ohashi}}, \bibinfo {author} {\bibfnamefont {F.}~\bibnamefont {Honda}},
  \bibinfo {author} {\bibfnamefont {T.}~\bibnamefont {Eto}}, \bibinfo {author}
  {\bibfnamefont {S.}~\bibnamefont {Kaji}}, \bibinfo {author} {\bibfnamefont
  {I.}~\bibnamefont {Minamitake}}, \bibinfo {author} {\bibfnamefont
  {G.}~\bibnamefont {Oomi}}, \bibinfo {author} {\bibfnamefont {S.}~\bibnamefont
  {Koiwai}}, \ and\ \bibinfo {author} {\bibfnamefont {Y.}~\bibnamefont
  {Uwatoko}},\ }\href@noop {} {\bibfield  {journal} {\bibinfo  {journal}
  {Physica B: Condensed Matter}\ }\textbf {\bibinfo {volume} {312-313}},\
  \bibinfo {pages} {443 } (\bibinfo {year} {2002})}\BibitemShut {NoStop}%
\bibitem [{\citenamefont {Braithwaite}\ \emph {et~al.}(2016)\citenamefont
  {Braithwaite}, \citenamefont {Knafo}, \citenamefont {Settai}, \citenamefont
  {Aoki}, \citenamefont {Kurahashi},\ and\ \citenamefont
  {Flouquet}}]{braithwaite16}%
  \BibitemOpen
  \bibfield  {author} {\bibinfo {author} {\bibfnamefont {D.}~\bibnamefont
  {Braithwaite}}, \bibinfo {author} {\bibfnamefont {W.}~\bibnamefont {Knafo}},
  \bibinfo {author} {\bibfnamefont {R.}~\bibnamefont {Settai}}, \bibinfo
  {author} {\bibfnamefont {D.}~\bibnamefont {Aoki}}, \bibinfo {author}
  {\bibfnamefont {S.}~\bibnamefont {Kurahashi}}, \ and\ \bibinfo {author}
  {\bibfnamefont {J.}~\bibnamefont {Flouquet}},\ }\href@noop {} {\bibfield
  {journal} {\bibinfo  {journal} {Review of Scientific Instruments}\ }\textbf
  {\bibinfo {volume} {87}},\ \bibinfo {eid} {023907} (\bibinfo {year}
  {2016})}\BibitemShut {NoStop}%
\bibitem [{\citenamefont {Settai}\ \emph {et~al.}(2015)\citenamefont {Settai},
  \citenamefont {Knafo}, \citenamefont {Braithwaite}, \citenamefont
  {Kurahashi}, \citenamefont {Aoki},\ and\ \citenamefont
  {Flouquet}}]{settai15}%
  \BibitemOpen
  \bibfield  {author} {\bibinfo {author} {\bibfnamefont {R.}~\bibnamefont
  {Settai}}, \bibinfo {author} {\bibfnamefont {W.}~\bibnamefont {Knafo}},
  \bibinfo {author} {\bibfnamefont {D.}~\bibnamefont {Braithwaite}}, \bibinfo
  {author} {\bibfnamefont {S.}~\bibnamefont {Kurahashi}}, \bibinfo {author}
  {\bibfnamefont {D.}~\bibnamefont {Aoki}}, \ and\ \bibinfo {author}
  {\bibfnamefont {J.}~\bibnamefont {Flouquet}},\ }\href@noop {} {\bibfield
  {journal} {\bibinfo  {journal} {Review of High Pressure Science and
  Technology / Koatsuryoku No Kagaku To Gijutsu}\ }\textbf {\bibinfo {volume}
  {25}},\ \bibinfo {pages} {325} (\bibinfo {year} {2015})}\BibitemShut
  {NoStop}%
\bibitem [{not({\natexlab{c}})}]{note_orb}%
  \BibitemOpen
  \bibinfo {note} {See Refs. \cite{scheerer12,scheerer14}, where a sample
  dependence of the orbital contribution to the high-field electrical
  resistivity, resulting from the cyclotron motion of conduction electrons, was
  evidenced in the heavy-fermion paramagnet URu$_2$Si$_2$.}\BibitemShut {Stop}%
\bibitem [{\citenamefont {\={O}nuki}\ \emph {et~al.}(2004)\citenamefont
  {\={O}nuki}, \citenamefont {Settai}, \citenamefont {Sugiyama}, \citenamefont
  {Takeuchi}, \citenamefont {Kobayashi}, \citenamefont {Haga},\ and\
  \citenamefont {Yamamoto}}]{onuki04}%
  \BibitemOpen
  \bibfield  {author} {\bibinfo {author} {\bibfnamefont {Y.}~\bibnamefont
  {\={O}nuki}}, \bibinfo {author} {\bibfnamefont {R.}~\bibnamefont {Settai}},
  \bibinfo {author} {\bibfnamefont {K.}~\bibnamefont {Sugiyama}}, \bibinfo
  {author} {\bibfnamefont {T.}~\bibnamefont {Takeuchi}}, \bibinfo {author}
  {\bibfnamefont {T.~C.}\ \bibnamefont {Kobayashi}}, \bibinfo {author}
  {\bibfnamefont {Y.}~\bibnamefont {Haga}}, \ and\ \bibinfo {author}
  {\bibfnamefont {E.}~\bibnamefont {Yamamoto}},\ }\href {\doibase
  10.1143/JPSJ.73.769} {\bibfield  {journal} {\bibinfo  {journal} {Journal of
  the Physical Society of Japan}\ }\textbf {\bibinfo {volume} {73}},\ \bibinfo
  {pages} {769} (\bibinfo {year} {2004})}\BibitemShut {NoStop}%
\bibitem [{\citenamefont {Flouquet}\ \emph {et~al.}(2010)\citenamefont
  {Flouquet}, \citenamefont {Aoki}, \citenamefont {Knafo}, \citenamefont
  {Knebel}, \citenamefont {Matsuda}, \citenamefont {Raymond}, \citenamefont
  {Proust}, \citenamefont {Paulsen},\ and\ \citenamefont {Haen}}]{flouquet10}%
  \BibitemOpen
  \bibfield  {author} {\bibinfo {author} {\bibfnamefont {J.}~\bibnamefont
  {Flouquet}}, \bibinfo {author} {\bibfnamefont {D.}~\bibnamefont {Aoki}},
  \bibinfo {author} {\bibfnamefont {W.}~\bibnamefont {Knafo}}, \bibinfo
  {author} {\bibfnamefont {G.}~\bibnamefont {Knebel}}, \bibinfo {author}
  {\bibfnamefont {T.~D.}\ \bibnamefont {Matsuda}}, \bibinfo {author}
  {\bibfnamefont {S.}~\bibnamefont {Raymond}}, \bibinfo {author} {\bibfnamefont
  {C.}~\bibnamefont {Proust}}, \bibinfo {author} {\bibfnamefont
  {C.}~\bibnamefont {Paulsen}}, \ and\ \bibinfo {author} {\bibfnamefont
  {P.}~\bibnamefont {Haen}},\ }\href@noop {} {\bibfield  {journal} {\bibinfo
  {journal} {Journal of Low Temperature Physics}\ }\textbf {\bibinfo {volume}
  {161}},\ \bibinfo {pages} {83} (\bibinfo {year} {2010})}\BibitemShut
  {NoStop}%
\bibitem [{\citenamefont {Aoki}\ \emph {et~al.}(2012)\citenamefont {Aoki},
  \citenamefont {Paulsen}, \citenamefont {Kotegawa}, \citenamefont {Hardy},
  \citenamefont {Meingast}, \citenamefont {Haen}, \citenamefont {Boukahil},
  \citenamefont {Knafo}, \citenamefont {Ressouche}, \citenamefont {Raymond},\
  and\ \citenamefont {Flouquet}}]{aoki12}%
  \BibitemOpen
  \bibfield  {author} {\bibinfo {author} {\bibfnamefont {D.}~\bibnamefont
  {Aoki}}, \bibinfo {author} {\bibfnamefont {C.}~\bibnamefont {Paulsen}},
  \bibinfo {author} {\bibfnamefont {H.}~\bibnamefont {Kotegawa}}, \bibinfo
  {author} {\bibfnamefont {F.}~\bibnamefont {Hardy}}, \bibinfo {author}
  {\bibfnamefont {C.}~\bibnamefont {Meingast}}, \bibinfo {author}
  {\bibfnamefont {P.}~\bibnamefont {Haen}}, \bibinfo {author} {\bibfnamefont
  {M.}~\bibnamefont {Boukahil}}, \bibinfo {author} {\bibfnamefont
  {W.}~\bibnamefont {Knafo}}, \bibinfo {author} {\bibfnamefont
  {E.}~\bibnamefont {Ressouche}}, \bibinfo {author} {\bibfnamefont
  {S.}~\bibnamefont {Raymond}}, \ and\ \bibinfo {author} {\bibfnamefont
  {J.}~\bibnamefont {Flouquet}},\ }\href {\doibase 10.1143/JPSJ.81.034711}
  {\bibfield  {journal} {\bibinfo  {journal} {Journal of the Physical Society
  of Japan}\ }\textbf {\bibinfo {volume} {81}},\ \bibinfo {pages} {034711}
  (\bibinfo {year} {2012})}\BibitemShut {NoStop}%
\bibitem [{\citenamefont {Haen}\ \emph {et~al.}(1996)\citenamefont {Haen},
  \citenamefont {Lapierre}, \citenamefont {Voiron},\ and\ \citenamefont
  {Fouquet}}]{haen96}%
  \BibitemOpen
  \bibfield  {author} {\bibinfo {author} {\bibfnamefont {P.}~\bibnamefont
  {Haen}}, \bibinfo {author} {\bibfnamefont {F.}~\bibnamefont {Lapierre}},
  \bibinfo {author} {\bibfnamefont {J.}~\bibnamefont {Voiron}}, \ and\ \bibinfo
  {author} {\bibfnamefont {J.}~\bibnamefont {Fouquet}},\ }\href@noop {}
  {\bibfield  {journal} {\bibinfo  {journal} {J. Phys. Soc. Jpn. (Suppl. B)}\
  }\textbf {\bibinfo {volume} {65}},\ \bibinfo {pages} {27} (\bibinfo {year}
  {1996})}\BibitemShut {NoStop}%
\bibitem [{\citenamefont {Aoki}\ \emph {et~al.}(2011)\citenamefont {Aoki},
  \citenamefont {Paulsen}, \citenamefont {Matsuda}, \citenamefont {Malone},
  \citenamefont {Knebel}, \citenamefont {Haen}, \citenamefont {Lejay},
  \citenamefont {Settai}, \citenamefont {Ōnuki},\ and\ \citenamefont
  {Flouquet}}]{aoki11}%
  \BibitemOpen
  \bibfield  {author} {\bibinfo {author} {\bibfnamefont {D.}~\bibnamefont
  {Aoki}}, \bibinfo {author} {\bibfnamefont {C.}~\bibnamefont {Paulsen}},
  \bibinfo {author} {\bibfnamefont {T.~D.}\ \bibnamefont {Matsuda}}, \bibinfo
  {author} {\bibfnamefont {L.}~\bibnamefont {Malone}}, \bibinfo {author}
  {\bibfnamefont {G.}~\bibnamefont {Knebel}}, \bibinfo {author} {\bibfnamefont
  {P.}~\bibnamefont {Haen}}, \bibinfo {author} {\bibfnamefont {P.}~\bibnamefont
  {Lejay}}, \bibinfo {author} {\bibfnamefont {R.}~\bibnamefont {Settai}},
  \bibinfo {author} {\bibfnamefont {Y.}~\bibnamefont {Ōnuki}}, \ and\ \bibinfo
  {author} {\bibfnamefont {J.}~\bibnamefont {Flouquet}},\ }\href {\doibase
  10.1143/JPSJ.80.053702} {\bibfield  {journal} {\bibinfo  {journal} {Journal
  of the Physical Society of Japan}\ }\textbf {\bibinfo {volume} {80}},\
  \bibinfo {pages} {053702} (\bibinfo {year} {2011})}\BibitemShut {NoStop}%
\bibitem [{\citenamefont {Pourret}\ \emph {et~al.}(2014)\citenamefont
  {Pourret}, \citenamefont {Aoki}, \citenamefont {Boukahil}, \citenamefont
  {Brison}, \citenamefont {Knafo}, \citenamefont {Knebel}, \citenamefont
  {Raymond}, \citenamefont {Taupin}, \citenamefont {Ōnuki},\ and\
  \citenamefont {Flouquet}}]{pourret14}%
  \BibitemOpen
  \bibfield  {author} {\bibinfo {author} {\bibfnamefont {A.}~\bibnamefont
  {Pourret}}, \bibinfo {author} {\bibfnamefont {D.}~\bibnamefont {Aoki}},
  \bibinfo {author} {\bibfnamefont {M.}~\bibnamefont {Boukahil}}, \bibinfo
  {author} {\bibfnamefont {J.-P.}\ \bibnamefont {Brison}}, \bibinfo {author}
  {\bibfnamefont {W.}~\bibnamefont {Knafo}}, \bibinfo {author} {\bibfnamefont
  {G.}~\bibnamefont {Knebel}}, \bibinfo {author} {\bibfnamefont
  {S.}~\bibnamefont {Raymond}}, \bibinfo {author} {\bibfnamefont
  {M.}~\bibnamefont {Taupin}}, \bibinfo {author} {\bibfnamefont
  {Y.}~\bibnamefont {Ōnuki}}, \ and\ \bibinfo {author} {\bibfnamefont
  {J.}~\bibnamefont {Flouquet}},\ }\href {\doibase 10.7566/JPSJ.83.061002}
  {\bibfield  {journal} {\bibinfo  {journal} {Journal of the Physical Society
  of Japan}\ }\textbf {\bibinfo {volume} {83}},\ \bibinfo {pages} {061002}
  (\bibinfo {year} {2014})}\BibitemShut {NoStop}%
\bibitem [{\citenamefont {Palacio~Morales}\ \emph {et~al.}(2015)\citenamefont
  {Palacio~Morales}, \citenamefont {Pourret}, \citenamefont {Seyfarth},
  \citenamefont {Suzuki}, \citenamefont {Braithwaite}, \citenamefont {Knebel},
  \citenamefont {Aoki},\ and\ \citenamefont {Flouquet}}]{palacio15}%
  \BibitemOpen
  \bibfield  {author} {\bibinfo {author} {\bibfnamefont {A.}~\bibnamefont
  {Palacio~Morales}}, \bibinfo {author} {\bibfnamefont {A.}~\bibnamefont
  {Pourret}}, \bibinfo {author} {\bibfnamefont {G.}~\bibnamefont {Seyfarth}},
  \bibinfo {author} {\bibfnamefont {M.-T.}\ \bibnamefont {Suzuki}}, \bibinfo
  {author} {\bibfnamefont {D.}~\bibnamefont {Braithwaite}}, \bibinfo {author}
  {\bibfnamefont {G.}~\bibnamefont {Knebel}}, \bibinfo {author} {\bibfnamefont
  {D.}~\bibnamefont {Aoki}}, \ and\ \bibinfo {author} {\bibfnamefont
  {J.}~\bibnamefont {Flouquet}},\ }\href@noop {} {\bibfield  {journal}
  {\bibinfo  {journal} {Phys. Rev. B}\ }\textbf {\bibinfo {volume} {91}},\
  \bibinfo {pages} {245129} (\bibinfo {year} {2015})}\BibitemShut {NoStop}%
\bibitem [{\citenamefont {Mori}\ \emph {et~al.}(1999)\citenamefont {Mori},
  \citenamefont {Takeshita}, \citenamefont {M\^{o}ri},\ and\ \citenamefont
  {Uwatoko}}]{mori99}%
  \BibitemOpen
  \bibfield  {author} {\bibinfo {author} {\bibfnamefont {H.}~\bibnamefont
  {Mori}}, \bibinfo {author} {\bibfnamefont {N.}~\bibnamefont {Takeshita}},
  \bibinfo {author} {\bibfnamefont {N.}~\bibnamefont {M\^{o}ri}}, \ and\
  \bibinfo {author} {\bibfnamefont {Y.}~\bibnamefont {Uwatoko}},\ }\href@noop
  {} {\bibfield  {journal} {\bibinfo  {journal} {Physica B: Condensed Matter}\
  }\textbf {\bibinfo {volume} {259-261}},\ \bibinfo {pages} {58 } (\bibinfo
  {year} {1999})}\BibitemShut {NoStop}%
\bibitem [{\citenamefont {Hanzawa}\ \emph {et~al.}(1985)\citenamefont
  {Hanzawa}, \citenamefont {Yamada},\ and\ \citenamefont {Yosida}}]{hanzawa85}%
  \BibitemOpen
  \bibfield  {author} {\bibinfo {author} {\bibfnamefont {K.}~\bibnamefont
  {Hanzawa}}, \bibinfo {author} {\bibfnamefont {K.}~\bibnamefont {Yamada}}, \
  and\ \bibinfo {author} {\bibfnamefont {K.}~\bibnamefont {Yosida}},\
  }\href@noop {} {\bibfield  {journal} {\bibinfo  {journal} {J. Magn. Magn.
  Mater.}\ }\textbf {\bibinfo {volume} {47-48}},\ \bibinfo {pages} {357}
  (\bibinfo {year} {1985})}\BibitemShut {NoStop}%
\bibitem [{\citenamefont {Watanabe}\ and\ \citenamefont
  {Miyake}(2011)}]{watanabe11}%
  \BibitemOpen
  \bibfield  {author} {\bibinfo {author} {\bibfnamefont {S.}~\bibnamefont
  {Watanabe}}\ and\ \bibinfo {author} {\bibfnamefont {K.}~\bibnamefont
  {Miyake}},\ }\href@noop {} {\bibfield  {journal} {\bibinfo  {journal}
  {Journal of Physics: Condensed Matter}\ }\textbf {\bibinfo {volume} {23}},\
  \bibinfo {pages} {094217} (\bibinfo {year} {2011})}\BibitemShut {NoStop}%
\bibitem [{\citenamefont {Matsuda}\ \emph {et~al.}(2014)\citenamefont
  {Matsuda}, \citenamefont {Her}, \citenamefont {Michimura}, \citenamefont
  {Inami}, \citenamefont {Ebihara},\ and\ \citenamefont
  {Amitsuka}}]{matsuda14}%
  \BibitemOpen
  \bibfield  {author} {\bibinfo {author} {\bibfnamefont {Y.~H.}\ \bibnamefont
  {Matsuda}}, \bibinfo {author} {\bibfnamefont {J.-L.}\ \bibnamefont {Her}},
  \bibinfo {author} {\bibfnamefont {S.}~\bibnamefont {Michimura}}, \bibinfo
  {author} {\bibfnamefont {T.}~\bibnamefont {Inami}}, \bibinfo {author}
  {\bibfnamefont {T.}~\bibnamefont {Ebihara}}, \ and\ \bibinfo {author}
  {\bibfnamefont {H.}~\bibnamefont {Amitsuka}},\ }\href@noop {} {\bibfield
  {journal} {\bibinfo  {journal} {JPS Conf. Proc.}\ }\textbf {\bibinfo {volume}
  {3}},\ \bibinfo {pages} {011044} (\bibinfo {year} {2014})}\BibitemShut
  {NoStop}%
\bibitem [{\citenamefont {Graf}\ \emph {et~al.}(1997)\citenamefont {Graf},
  \citenamefont {Thompson}, \citenamefont {Hundley}, \citenamefont
  {Movshovich}, \citenamefont {Fisk}, \citenamefont {Mandrus}, \citenamefont
  {Fisher},\ and\ \citenamefont {Phillips}}]{graf97}%
  \BibitemOpen
  \bibfield  {author} {\bibinfo {author} {\bibfnamefont {T.}~\bibnamefont
  {Graf}}, \bibinfo {author} {\bibfnamefont {J.~D.}\ \bibnamefont {Thompson}},
  \bibinfo {author} {\bibfnamefont {M.~F.}\ \bibnamefont {Hundley}}, \bibinfo
  {author} {\bibfnamefont {R.}~\bibnamefont {Movshovich}}, \bibinfo {author}
  {\bibfnamefont {Z.}~\bibnamefont {Fisk}}, \bibinfo {author} {\bibfnamefont
  {D.}~\bibnamefont {Mandrus}}, \bibinfo {author} {\bibfnamefont {R.~A.}\
  \bibnamefont {Fisher}}, \ and\ \bibinfo {author} {\bibfnamefont {N.~E.}\
  \bibnamefont {Phillips}},\ }\href@noop {} {\bibfield  {journal} {\bibinfo
  {journal} {Phys. Rev. Lett.}\ }\textbf {\bibinfo {volume} {78}},\ \bibinfo
  {pages} {3769} (\bibinfo {year} {1997})}\BibitemShut {NoStop}%
\bibitem [{\citenamefont {Lacerda}\ \emph {et~al.}(1989)\citenamefont
  {Lacerda}, \citenamefont {de~Visser}, \citenamefont {Haen}, \citenamefont
  {Lejay},\ and\ \citenamefont {Flouquet}}]{lacerda89}%
  \BibitemOpen
  \bibfield  {author} {\bibinfo {author} {\bibfnamefont {A.}~\bibnamefont
  {Lacerda}}, \bibinfo {author} {\bibfnamefont {A.}~\bibnamefont {de~Visser}},
  \bibinfo {author} {\bibfnamefont {P.}~\bibnamefont {Haen}}, \bibinfo {author}
  {\bibfnamefont {P.}~\bibnamefont {Lejay}}, \ and\ \bibinfo {author}
  {\bibfnamefont {J.}~\bibnamefont {Flouquet}},\ }\href@noop {} {\bibfield
  {journal} {\bibinfo  {journal} {Phys. Rev. B}\ }\textbf {\bibinfo {volume}
  {40}},\ \bibinfo {pages} {8759} (\bibinfo {year} {1989})}\BibitemShut
  {NoStop}%
\bibitem [{not({\natexlab{d}})}]{note}%
  \BibitemOpen
  \bibinfo {note} {The Gr\"{u}neisen parameter $\Gamma=7$ is calculated in the
  pressure range $p>p_c$ following $\Gamma=\partial\gamma/\partial
  p/(\gamma\kappa)$, where $\partial\gamma/\partial p$ is extracted from
  \cite{graf97} and the compressibility coefficient $\kappa=0.74$~Mbar$^{-1}$
  is extracted from \cite{ohashi03}.}\BibitemShut {Stop}%
\bibitem [{\citenamefont {Araki}\ \emph {et~al.}(2001)\citenamefont {Araki},
  \citenamefont {Settai}, \citenamefont {Kobayashi}, \citenamefont {Harima},\
  and\ \citenamefont {\ifmmode~\bar{O}\else \={O}\fi{}nuki}}]{araki01}%
  \BibitemOpen
  \bibfield  {author} {\bibinfo {author} {\bibfnamefont {S.}~\bibnamefont
  {Araki}}, \bibinfo {author} {\bibfnamefont {R.}~\bibnamefont {Settai}},
  \bibinfo {author} {\bibfnamefont {T.~C.}\ \bibnamefont {Kobayashi}}, \bibinfo
  {author} {\bibfnamefont {H.}~\bibnamefont {Harima}}, \ and\ \bibinfo {author}
  {\bibfnamefont {Y.}~\bibnamefont {\ifmmode~\bar{O}\else \={O}\fi{}nuki}},\
  }\href@noop {} {\bibfield  {journal} {\bibinfo  {journal} {Phys. Rev. B}\
  }\textbf {\bibinfo {volume} {64}},\ \bibinfo {pages} {224417} (\bibinfo
  {year} {2001})}\BibitemShut {NoStop}%
\bibitem [{\citenamefont {Sheikin}()}]{sheikinxx}%
  \BibitemOpen
  \bibfield  {author} {\bibinfo {author} {\bibfnamefont {I.}~\bibnamefont
  {Sheikin}},\ }\href@noop {} {\bibinfo  {journal} {(unpublished)}\
  }\BibitemShut {NoStop}%
\bibitem [{dej(1990)}]{dejongh90}%
  \BibitemOpen
\bibfield  {journal} {  }\href@noop {} {\emph {\bibinfo {title} {Magnetic
  properties of layered transition metal compounds}}}\ (\bibinfo  {publisher}
  {Kluwer Academic Publishers, Dordrecht / Boston / London, Edited by L.J. De
  Jongh},\ \bibinfo {year} {1990})\BibitemShut {NoStop}%
\bibitem [{\citenamefont {Johnston}(1989)}]{johnston89}%
  \BibitemOpen
  \bibfield  {author} {\bibinfo {author} {\bibfnamefont {D.~C.}\ \bibnamefont
  {Johnston}},\ }\href@noop {} {\bibfield  {journal} {\bibinfo  {journal}
  {Phys. Rev. Lett.}\ }\textbf {\bibinfo {volume} {62}},\ \bibinfo {pages}
  {957} (\bibinfo {year} {1989})}\BibitemShut {NoStop}%
\bibitem [{\citenamefont {Pines}(2013)}]{pines13}%
  \BibitemOpen
  \bibfield  {author} {\bibinfo {author} {\bibfnamefont {D.}~\bibnamefont
  {Pines}},\ }\href {\doibase 10.1021/jp403088e} {\bibfield  {journal}
  {\bibinfo  {journal} {The Journal of Physical Chemistry B}\ }\textbf
  {\bibinfo {volume} {117}},\ \bibinfo {pages} {13145} (\bibinfo {year}
  {2013})}\BibitemShut {NoStop}%
\bibitem [{\citenamefont {Scheerer}\ \emph {et~al.}(2014)\citenamefont
  {Scheerer}, \citenamefont {Knafo}, \citenamefont {Aoki}, \citenamefont
  {Nardone}, \citenamefont {Zitouni}, \citenamefont {B\'eard}, \citenamefont
  {Billette}, \citenamefont {Barata}, \citenamefont {Jaudet}, \citenamefont
  {Suleiman}, \citenamefont {Frings}, \citenamefont {Drigo}, \citenamefont
  {Audouard}, \citenamefont {Matsuda}, \citenamefont {Pourret}, \citenamefont
  {Knebel},\ and\ \citenamefont {Flouquet}}]{scheerer14}%
  \BibitemOpen
  \bibfield  {author} {\bibinfo {author} {\bibfnamefont {G.~W.}\ \bibnamefont
  {Scheerer}}, \bibinfo {author} {\bibfnamefont {W.}~\bibnamefont {Knafo}},
  \bibinfo {author} {\bibfnamefont {D.}~\bibnamefont {Aoki}}, \bibinfo {author}
  {\bibfnamefont {M.}~\bibnamefont {Nardone}}, \bibinfo {author} {\bibfnamefont
  {A.}~\bibnamefont {Zitouni}}, \bibinfo {author} {\bibfnamefont
  {J.}~\bibnamefont {B\'eard}}, \bibinfo {author} {\bibfnamefont
  {J.}~\bibnamefont {Billette}}, \bibinfo {author} {\bibfnamefont
  {J.}~\bibnamefont {Barata}}, \bibinfo {author} {\bibfnamefont
  {C.}~\bibnamefont {Jaudet}}, \bibinfo {author} {\bibfnamefont
  {M.}~\bibnamefont {Suleiman}}, \bibinfo {author} {\bibfnamefont
  {P.}~\bibnamefont {Frings}}, \bibinfo {author} {\bibfnamefont
  {L.}~\bibnamefont {Drigo}}, \bibinfo {author} {\bibfnamefont
  {A.}~\bibnamefont {Audouard}}, \bibinfo {author} {\bibfnamefont {T.~D.}\
  \bibnamefont {Matsuda}}, \bibinfo {author} {\bibfnamefont {A.}~\bibnamefont
  {Pourret}}, \bibinfo {author} {\bibfnamefont {G.}~\bibnamefont {Knebel}}, \
  and\ \bibinfo {author} {\bibfnamefont {J.}~\bibnamefont {Flouquet}},\
  }\href@noop {} {\bibfield  {journal} {\bibinfo  {journal} {Phys. Rev. B}\
  }\textbf {\bibinfo {volume} {89}},\ \bibinfo {pages} {165107} (\bibinfo
  {year} {2014})}\BibitemShut {NoStop}%
\end{thebibliography}
\end{document}